\documentclass[twocolumn,english]{revtex4-1}
\usepackage{mathptmx}
\usepackage{graphicx}
\usepackage[T1]{fontenc}
\usepackage[latin9]{inputenc}
\usepackage{color}
\usepackage{booktabs}
\usepackage{mathtools}
\usepackage{amsmath}
\usepackage{amssymb}
\usepackage{stmaryrd}
\usepackage{esint}
\usepackage{upgreek}
\usepackage[unicode=true,pdfusetitle,
 bookmarks=true,bookmarksnumbered=false,bookmarksopen=false,
 breaklinks=false,pdfborder={0 0 1},backref=false,colorlinks=true]
 {hyperref}
\hypersetup{
 linkcolor=blue,urlcolor={blue},filecolor={blue},citecolor={blue}}

\makeatletter

\providecommand{\tabularnewline}{\\}

\usepackage{babel}
\usepackage{amsopn}\usepackage{amsthm}\usepackage{amscd}\usepackage{amsfonts}\usepackage{chngcntr}

\usepackage{times}

\counterwithout{figure}{section}
\counterwithout{equation}{section}
\usepackage{changepage}

\counterwithout{figure}{section}
\counterwithout{equation}{section}
\usepackage{times}
\usepackage{bm}
\usepackage[caption=false]{subfig}
\usepackage{floatrow}

\renewcommand{\fnum@figure}{FIG. \thefigure}

\makeatother

\begin{document}
\global\long\def\real{\mathbb{R}}%

\global\long\def\RefVol{\Omega_{0}}%

\global\long\def\Refx{\mathbf{X}}%

\global\long\def\Curx{\mathbf{x}}%

\global\long\def\map{\boldsymbol{\chi}}%

\global\long\def\defgrad{\mathbf{F}}%

\global\long\def\defgradT{\mathbf{F}^{\mathrm{T}}}%

\global\long\def\defgradTi{\mathbf{F}^{\mathrm{-T}}}%

\global\long\def\d{\mathrm{d}}%

\global\long\def\RCG{\mathbf{C}}%

\global\long\def\LCG{\mathbf{b}}%

\global\long\def\div#1{\nabla\cdot#1}%

\global\long\def\curl#1{\nabla\times#1}%

\global\long\def\T#1{#1^{\mathrm{T}}}%

\global\long\def\CurStress{\boldsymbol{\sigma}}%

\global\long\def\Refcurl#1{\nabla_{\mathbf{X}}\times#1}%

\global\long\def\s#1{#1^{\star}}%

\global\long\def\m#1{#1^{(m)}}%

\global\long\def\f#1{#1^{(f)}}%

\global\long\def\p#1{#1^{(p)}}%

\global\long\def\Refdiv#1{\nabla_{\mathbf{X}}\cdot#1}%

\global\long\def\phase#1{#1^{\left(p\right)}}%

\global\long\def\jm{j_{m}^{(p)}}%

\global\long\def\xinc{\mathbf{\dot{x}}}%

\global\long\def\Pinc{\mathbf{\dot{P}}}%

\global\long\def\Einc{\mathbf{\dot{E}}}%

\global\long\def\Finc{\mathbf{\dot{F}}}%

\global\long\def\dinc{\mathbf{\check{d}}}%

\global\long\def\einc{\mathbf{\check{e}}}%

\global\long\def\Dinc{\mathbf{\dot{D}}}%

\global\long\def\sinc{\boldsymbol{\Sigma}}%

\global\long\def\fA{\mathbb{\boldsymbol{\mathscr{A}}}}%

\global\long\def\fB{\mathbb{\boldsymbol{\mathscr{B}}}}%

\global\long\def\fC{\mathbb{\boldsymbol{\mathscr{C}}}}%

\global\long\def\plane{\left(x_{1},x_{3}\right)}%

\global\long\def\anti{\dot{x}_{2}}%

\global\long\def\bnd{\partial\Omega}%

\global\long\def\rcg{\mathbf{C}}%

\global\long\def\lcg{\mathbf{b}}%

\global\long\def\rcgcomp#1{C_{#1}}%

\global\long\def\cronck#1{\delta_{#1}}%

\global\long\def\lcgcomp#1{b_{#1}}%

\global\long\def\deformation{\boldsymbol{\chi}}%

\global\long\def\cauchycomp#1{\sigma_{#1}}%

\global\long\def\idgt{\dg^{\mathrm{-T}}}%

\global\long\def\cauchy{\boldsymbol{\sigma}}%

\global\long\def\normal{\mathbf{n}}%

\global\long\def\normall{\mathbf{N}}%

\global\long\def\traction{\mathbf{t}}%

\global\long\def\tractionl{\mathbf{t}_{L}}%

\global\long\def\ed{\mathbf{d}}%

\global\long\def\edcomp#1{d_{#1}}%

\global\long\def\edl{\mathbf{D}}%

\global\long\def\edlcomp#1{D_{#1}}%

\global\long\def\ef{\mathbf{e}}%

\global\long\def\efcomp#1{e_{#1}}%

\global\long\def\efl{\mathbf{E}}%

\global\long\def\freech{q_{e}}%

\global\long\def\surfacech{w_{e}}%

\global\long\def\outer#1{#1^{\star}}%

\global\long\def\perm{\epsilon_{0}}%

\global\long\def\matper{\epsilon}%

\global\long\def\jump#1{\llbracket#1\rrbracket}%

\global\long\def\identity{\mathbf{I}}%

\global\long\def\area{\mathrm{d}a}%

\global\long\def\areal{\mathrm{d}A}%

\global\long\def\refsys{\mathbf{X}}%

\global\long\def\Grad{\nabla_{\refsys}}%

\global\long\def\grad{\nabla}%

\global\long\def\divg{\nabla\cdot}%

\global\long\def\Div{\nabla_{\refsys}}%

\global\long\def\derivative#1#2{\frac{\partial#1}{\partial#2}}%

\global\long\def\aef{\Psi}%

\global\long\def\dltendl{\edl\otimes\edl}%

\global\long\def\tr#1{\mathrm{tr}#1}%

\global\long\def\ii#1{I_{#1}}%

\global\long\def\dh{\hat{D}}%

\global\long\def\inc#1{\dot{#1}}%

\global\long\def\sys{\mathbf{x}}%

\global\long\def\curl{\nabla}%

\global\long\def\Curl{\nabla_{\refsys}}%

\global\long\def\piolaincpush{\boldsymbol{\Sigma}}%

\global\long\def\piolaincpushcomp#1{\Sigma_{#1}}%

\global\long\def\elaspush{\boldsymbol{\mathcal{C}}}%

\global\long\def\elecpush{\boldsymbol{\mathcal{A}}}%

\global\long\def\elaselecpush{\boldsymbol{\mathcal{B}}}%

\global\long\def\disgrad{\mathbf{h}}%

\global\long\def\disgradcomp#1{h_{#1}}%

\global\long\def\trans#1{#1^{\mathrm{\mathsf{T}}}}%

\global\long\def\phase#1{#1^{\left(p\right)}}%

\global\long\def\elecpushcomp#1{\mathcal{A}_{#1}}%

\global\long\def\elaselecpushcomp#1{\mathcal{B}_{#1}}%

\global\long\def\elaspushcomp#1{\mathcal{C}_{#1}}%

\global\long\def\dnh{\aef_{DH}}%

\global\long\def\woo{\varsigma}%

\global\long\def\wif{\Lambda}%

\global\long\def\structurefun{S}%

\global\long\def\dg{\mathbf{F}}%

\global\long\def\dgcomp#1{F_{#1}}%

\global\long\def\piola{\mathbf{P}}%

\global\long\def\refbody{\Omega_{0}}%

\global\long\def\refbnd{\partial\refbody}%

\global\long\def\bnd{\partial\Omega}%

\global\long\def\rcg{\mathbf{C}}%

\global\long\def\lcg{\mathbf{b}}%

\global\long\def\rcgcomp#1{C_{#1}}%

\global\long\def\cronck#1{\delta_{#1}}%

\global\long\def\lcgcomp#1{b_{#1}}%

\global\long\def\deformation{\boldsymbol{\chi}}%

\global\long\def\dgt{\dg^{\mathrm{T}}}%

\global\long\def\idgcomp#1{F_{#1}^{-1}}%

\global\long\def\velocity{\mathbf{v}}%

\global\long\def\accel{\mathbf{a}}%

\global\long\def\vg{\mathbf{l}}%

\global\long\def\idg{\dg^{-1}}%

\global\long\def\cauchycomp#1{\sigma_{#1}}%

\global\long\def\idgt{\dg^{\mathrm{-T}}}%

\global\long\def\cauchy{\boldsymbol{\sigma}}%

\global\long\def\normal{\mathbf{n}}%

\global\long\def\normall{\mathbf{N}}%

\global\long\def\traction{\mathbf{t}}%

\global\long\def\tractionl{\mathbf{t}_{L}}%

\global\long\def\ed{\mathbf{d}}%

\global\long\def\edcomp#1{d_{#1}}%

\global\long\def\edl{\mathbf{D}}%

\global\long\def\edlcomp#1{D_{#1}}%

\global\long\def\ef{\mathbf{e}}%

\global\long\def\efcomp#1{e_{#1}}%

\global\long\def\efl{\mathbf{E}}%

\global\long\def\freech{q_{e}}%

\global\long\def\surfacech{w_{e}}%

\global\long\def\outer#1{#1^{\star}}%

\global\long\def\perm{\epsilon_{0}}%

\global\long\def\matper{\epsilon}%

\global\long\def\jump#1{\llbracket#1\rrbracket}%

\global\long\def\identity{\mathbf{I}}%

\global\long\def\area{\mathrm{d}a}%

\global\long\def\areal{\mathrm{d}A}%

\global\long\def\refsys{\mathbf{X}}%

\global\long\def\Grad{\nabla_{\refsys}}%

\global\long\def\grad{\nabla}%

\global\long\def\divg{\nabla\cdot}%

\global\long\def\Div{\nabla_{\refsys}}%

\global\long\def\derivative#1#2{\frac{\partial#1}{\partial#2}}%

\global\long\def\aef{\Psi}%

\global\long\def\dltendl{\edl\otimes\edl}%

\global\long\def\tr#1{\mathrm{tr}\left(#1\right)}%

\global\long\def\ii#1{I_{#1}}%

\global\long\def\dh{\hat{D}}%

\global\long\def\lage{\mathbf{E}}%

\global\long\def\inc#1{\dot{#1}}%

\global\long\def\sys{\mathbf{x}}%

\global\long\def\curl{\nabla}%

\global\long\def\Curl{\nabla_{\refsys}}%

\global\long\def\piolaincpush{\boldsymbol{\Sigma}}%

\global\long\def\piolaincpushcomp#1{\Sigma_{#1}}%

\global\long\def\edlincpush{\check{\mathbf{d}}}%

\global\long\def\edlincpushcomp#1{\check{d}_{#1}}%

\global\long\def\efincpush{\check{\mathbf{e}}}%

\global\long\def\efincpushcomp#1{\check{e}_{#1}}%

\global\long\def\elaspush{\boldsymbol{\mathcal{C}}}%

\global\long\def\elecpush{\boldsymbol{\mathcal{A}}}%

\global\long\def\elaselecpush{\boldsymbol{\mathcal{B}}}%

\global\long\def\disgrad{\mathbf{h}}%

\global\long\def\disgradcomp#1{h_{#1}}%

\global\long\def\trans#1{#1^{\mathrm{T}}}%

\global\long\def\phase#1{#1^{\left(p\right)}}%

\global\long\def\elecpushcomp#1{\mathcal{A}_{#1}}%

\global\long\def\elaselecpushcomp#1{\mathcal{B}_{#1}}%

\global\long\def\elaspushcomp#1{\mathcal{C}_{#1}}%

\global\long\def\dnh{\aef_{DG}}%

\global\long\def\dnhc{\mu\lambda^{2}}%

\global\long\def\dnhcc{\frac{\mu}{\lambda^{2}}+\frac{1}{\matper}d_{2}^{2}}%

\global\long\def\dnhb{\frac{1}{\matper}d_{2}}%

\global\long\def\afreq{\omega}%

\global\long\def\dispot{\phi}%

\global\long\def\edpot{\varphi}%

\global\long\def\kh{\hat{k}}%

\global\long\def\afreqh{\hat{\afreq}}%

\global\long\def\phasespeed{c}%

\global\long\def\bulkspeed{c_{B}}%

\global\long\def\speedh{\hat{c}}%

\global\long\def\dhth{\dh_{th}}%

\global\long\def\bulkspeedl{\bulkspeed_{\lambda}}%

\global\long\def\khth{\hat{k}_{th}}%

\global\long\def\p#1{#1^{\left(p\right)}}%

\global\long\def\maxinccomp#1{\inc{\outer{\sigma}}_{#1}}%

\global\long\def\maxcomp#1{\outer{\sigma}_{#1}}%

\global\long\def\relper{\matper_{r}}%

\global\long\def\sdh{\hat{d}}%

\global\long\def\iee{\varphi}%

\global\long\def\effectivemu{\tilde{\mu}}%

\global\long\def\fb#1{#1^{\left(a\right)}}%

\global\long\def\mt#1{#1^{\left(b\right)}}%

\global\long\def\phs#1{#1^{\left(p\right)}}%

\global\long\def\thc{h}%

\global\long\def\state{\mathbf{s}}%

\global\long\def\harmonicper{\breve{\matper}}%

\global\long\def\kb{k_{B}}%

\global\long\def\cb{\bar{c}}%

\global\long\def\mb{\bar{\mu}}%

\global\long\def\rb{\bar{\rho}}%

\global\long\def\wavenumber{k}%

\global\long\def\nh{\mathbf{n}}%

\global\long\def\mh{\mathbf{m}}%

\global\long\def\deflect{\inc x_{2}}%

\global\long\def\sdd#1{#1_{2,11}}%

\global\long\def\sdddd#1{#1_{2,1111}}%

\global\long\def\sd#1{#1_{2,1}}%

\global\long\def\sddd#1{#1_{2,111}}%

\global\long\def\xdddd#1{#1_{,\xi\xi\xi\xi}}%

\global\long\def\xdd#1{#1_{,\xi\xi}}%

\global\long\def\xd#1{#1_{,\xi}}%

\global\long\def\xddd#1{#1_{,\xi\xi\xi}}%

\global\long\def\jm{J_{m}}%

\global\long\def\dv{\Delta V}%

\global\long\def\ih{\mathbf{i}_{1}}%

\global\long\def\kh{\mathbf{i}_{3}}%

\global\long\def\jh{\mathbf{i}_{2}}%

\global\long\def\etil{E}%

\global\long\def\genT{\mathsf{Q}}%

\global\long\def\transfer{\mathsf{T}}%

\global\long\def\statevec{\mathbf{s}}%

\global\long\def\coefvec{\mathbf{c}}%

\global\long\def\pressure{p_{0}}%

\global\long\def\ncell#1{#1_{\left(n\right)}}%

\global\long\def\ydisp{\inc x_{2}}%

\global\long\def\ycord{x_{2}}%

\global\long\def\pn#1{\ncell{#1}^{\left(p\right)}}%

\global\long\def\pnm#1{#1_{\left(n\right)m}^{\left(p\right)}}%

\global\long\def\eigen{\boldsymbol{\eta}}%

\global\long\def\xcomp{x_{1}}%

\global\long\def\totalT{\mathsf{T_{\mathrm{tot}}}}%

\global\long\def\rads{\frac{\mathrm{rad}}{\mathrm{s}}}%

\global\long\def\lf{\gamma}%

\global\long\def\tf{T_{m}}%

\global\long\def\eigenim{\beta}%

\global\long\def\bS{\mathsf{S}}%

\global\long\def\Blochwn{{k}_{\mathit{B}}}%

\global\long\def\Blochwnx{k_{B1}}%

\global\long\def\Blochwny{k_{B2}}%

\global\long\def\waven{k}%

\global\long\def\Beff{\tilde{B}}%

\global\long\def\rhoaux{\check{\rho}}%

\global\long\def\nuaux{\check{\nu}}%

\global\long\def\Daux{\check{D}}%

\global\long\def\geff{\tilde{g}}%

\global\long\def\avevarsigma{\overline{\varsigma}}%

\global\long\def\Deff{\tilde{D}}%

\global\long\def\Dxeff{\tilde{D}_{1}}%

\global\long\def\Dyeff{\tilde{D}_{2}}%

\global\long\def\Dnueff{\tilde{D}_{\nu}}%

\global\long\def\Dxyeff{\tilde{D}_{12}}%

\global\long\def\nueff{\tilde{\nu}}%

\global\long\def\rhoeff{\tilde{\rho}}%

\global\long\def\aveR{\overline{R}}%

\global\long\def\aveMx{\overline{M}_{1}}%

\global\long\def\aveMy{\overline{M}_{2}}%

\global\long\def\aveMxy{\overline{M}_{12}}%

\global\long\def\aveM{\overline{M}}%

\global\long\def\avew{\overline{w}}%

\global\long\def\avep{\overline{p}}%

\global\long\def\vectora{\boldsymbol{\psi}}%

\global\long\def\vectorb{\boldsymbol{\varphi}}%
\global\long\def\avevectora{\overline{\boldsymbol{\psi}} }%

\global\long\def\avevectorao{\left\langle \boldsymbol{\psi}_{0}\right\rangle }%

\global\long\def\avevectorb{\overline{\boldsymbol{\varphi}} }%

\global\long\def\avevectorp{\overline{\mathsf{P}} }%
\global\long\def\DDeff{\tilde{\tilde{D}}}%
\global\long\def\DDxeff{\tilde{\tilde{D}}_{1}}%

\global\long\def\DDyeff{\tilde{\tilde{D}}_{2}}%

\global\long\def\DDnueff{\tilde{\tilde{D}}_{\nu}}%

\global\long\def\nuueff{\tilde{\tilde{\nu}}}%

\global\long\def\DDkeff{\tilde{\tilde{D}}_{k}}%

\global\long\def\DDxyeff{\tilde{\tilde{D}}_{12}}%

\global\long\def\rhooeff{\tilde{\tilde{\rho}}}%

\global\long\def\angRR{\left\langle \left\langle R_{\mathrm{p}}\right\rangle \right\rangle }%

\global\long\def\angMMx{\left\langle M_{1\mathrm{p}}\right\rangle }%

\global\long\def\angMMy{\left\langle M_{2\mathrm{p}}\right\rangle }%

\global\long\def\angMMxy{\left\langle M_{12\mathrm{p}}\right\rangle }%

\global\long\def\angww{\left\langle w_{\mathrm{p}}\right\rangle }%

\global\long\def\angpp{ \left\langle p_{\mathrm{p}}\right\rangle }%

\global\long\def\uF{u_{\mathrm{F}}}%

\global\long\def\wF{w_{\mathrm{F}}}%

\global\long\def\rhoF{\rho_{\mathrm{F}}}%

\global\long\def\BF{B_{\mathrm{F}}}%

\global\long\def\dl{l^{(i)}}%

\global\long\def\bF{\state_{\mathrm{d}}}%

\global\long\def\bA{\state_{\mathrm{f}}}%

\global\long\def\ba{\coefvec}%

\global\long\def\bcero{\mathsf{0}}%

\global\long\def\bH{\mathsf{H}}%

\global\long\def\ltotal{l_{\mathrm{tot}}}%

\global\long\def\bI{\mathsf{I}}%

\global\long\def\qpd{\genT_{\mathrm{d}}^{+}}%

\global\long\def\qmd{\genT_{\mathrm{d}}^{-}}%

\global\long\def\qpf{\genT_{\mathrm{f}}^{+}}%

\global\long\def\qmf{\genT_{\mathrm{f}}^{-}}%

\global\long\def\curcon{\Omega}%

\global\long\def\ld{\mathbf{D}}%

\global\long\def\qpdz{\genT_{\mathrm{d}\left(0\right)}^{+}}%

\global\long\def\qmdz{\genT_{\mathrm{d}\left(0\right)}^{-}}%

\global\long\def\qpfz{\genT_{\mathrm{f}\left(0\right)}^{+}}%

\global\long\def\qmfz{\genT_{\mathrm{f}\left(0\right)}^{-}}%

\global\long\def\qpdl{\genT_{\mathrm{d}\left(M\right)}^{+}}%

\global\long\def\qmdl{\genT_{\mathrm{d}\left(M\right)}^{-}}%

\global\long\def\qpfl{\genT_{\mathrm{f}\left(M\right)}^{+}}%

\global\long\def\qmfl{\genT_{\mathrm{f}\left(M\right)}^{-}}%

\global\long\def\e{\mathop{{\rm \mbox{{\Large e}}}}\nolimits}%

\global\long\def\Tr{\textrm{Tr}}%

\global\long\def\Det{\textrm{Det}}%

\global\long\def\sgn{\textrm{sgn}}%

\global\long\def\pr{^{\prime}}%

\global\long\def\bn#1{\mbox{\boldmath\ensuremath{#1}}}%

\global\long\def\bB{\bn B}%

\global\long\def\bP{\bn P}%

\global\long\def\bY{\bn Y}%

\global\long\def\bV{\bn V}%

\global\long\def\bW{\bn W}%

\global\long\def\bG{\bn G}%

\global\long\def\bM{\bn M}%

\global\long\def\bm{\bn m}%

\global\long\def\bE{\bn E}%

\global\long\def\bK{\bn K}%

\global\long\def\bL{\bn L}%

\global\long\def\bC{\bn C}%

\global\long\def\bT{\bn T}%

\global\long\def\bg{\bn g}%

\global\long\def\bN{\bn N}%

\global\long\def\bX{\bn X}%

\global\long\def\bR{\bn R}%

\global\long\def\bHs{\bn{Hs}}%

\global\long\def\bDelta{\bn{\Delta}}%

\global\long\def\bsigma{\bn{\sigma}}%

\global\long\def\bpsi{\bn{\psi}}%

\global\long\def\bQ{\bn Q}%

\global\long\def\bZ{\bn Z}%

\global\long\def\bU{\bn U}%

\global\long\def\bz{\bn z}%

\global\long\def\bu{\bn u}%

\global\long\def\bk{\bn k}%

\global\long\def\bUpsilon{\bn{\Upsilon}}%

\global\long\def\bEta{\bn{\eta}}%

\global\long\def\bmu{\bn{\mu}}%

\global\long\def\brho{\bn{\rho}}%

\global\long\def\bpartial{\bn{\partial}}%

\global\long\def\Elecf{\mathbf{E}}%

\global\long\def\Edisplacement{\mathbf{D}}%

\global\long\def\Strain{\grad\disp}%

\global\long\def\Stress{\boldsymbol{\sigma}}%

\global\long\def\Piezoelectricmodule{\mathbf{B}}%

\global\long\def\Permittivity{\mathbf{A}}%

\global\long\def\alfacoupled{\check{\alpha}}%

\global\long\def\gammacoupled{\check{\gamma}}%

\global\long\def\bettacoupled{\check{\beta}}%

\global\long\def\rhocoupled{\check{\rho}}%

\global\long\def\Scoupled{\check{S}}%

\global\long\def\Lcoupled{\check{L}}%

\global\long\def\Qcoupled{\check{Q}}%

\global\long\def\elas{\mathbf{C}}%

\global\long\def\momentum{\mathbf{p}}%

\global\long\def\force{\mathbf{f}}%

\global\long\def\disp{\mathbf{u}}%

\global\long\def\potential{\phi}%

\global\long\def\state{\mathsf{S}}%

\global\long\def\position{\mathbf{x}}%

\global\long\def\cmat{\mathsf{C}}%

\global\long\def\rmat{\mathsf{R}}%

\global\long\def\mmat{\mathsf{m}}%

\global\long\def\pmat{\mathsf{p}}%

\global\long\def\gradt{\mathsf{B}}%

\global\long\def\genpotential{\mathsf{w}}%

\global\long\def\vol{\Omega}%

\global\long\def\rve{\Omega^{\mathrm{rve}}}%

\global\long\def\effective#1{\tilde{#1}}%
\global\long\def\ensemble#1{\left\langle #1\right\rangle }%
\global\long\def\average#1{\overline{#1}}%
\global\long\def\rg{\mathbf{W}}%
\global\long\def\po#1{#1^{\left(1\right)}}%

\global\long\def\pt#1{#1^{\left(2\right)}}%

\global\long\def\ptr#1{#1^{\left(3\right)}}%

\global\long\def\length{l}%
\global\long\def\dt#1{\dot{#1}}%
\global\long\def\willis{\mathbf{S}}%
\global\long\def\area{a}%
\global\long\def\fmat{\mathsf{f}}%
\global\long\def\lmat{\mathsf{L}}%
\global\long\def\hmat{\mathsf{h}}%
\global\long\def\bmat{\mathsf{b}}%

\global\long\def\green{\mathsf{G}}%
\global\long\def\adjoint#1{#1^{\dagger}}%
\global\long\def\cc#1{#1^{*}}%
\global\long\def\intvol#1{\int_{\vol}#1\mathrm{d}\vol}%
\global\long\def\boundaryw{\bnd_{\mathsf{w}}}%
\global\long\def\intsurfacepot#1{\int_{\boundaryw}#1\mathrm{d}\area}%
\global\long\def\boundaryt{\bnd_{\mathsf{t}}}%
\global\long\def\intsurfaceh#1{\int_{\boundaryt}#1\mathrm{d}\area}%
\global\long\def\normalop{\mathsf{n}}%
\global\long\def\smat{\mathsf{s}}%
\global\long\def\fmat{\mathsf{f}}%
\global\long\def\dmat{\mathsf{D}}%
\global\long\def\zm#1#2{\mathsf{0}_{#1\times#2}}%
\global\long\def\nmat{\mathsf{N}}%
\global\long\def\tp#1{#1^{\mathsf{T}}}%
\global\long\def\charge{q}%
\global\long\def\scharge{\omega_{\mathrm{c}}}%
\global\long\def\dirac{\delta}%
\global\long\def\period{l}%
\global\long\def\scalarf{f}%
\global\long\def\scalarD{D}%
\global\long\def\randomvar{p}%
\global\long\def\cell{\vol_{\mathrm{p}}}%
\global\long\def\realization{y}%
\global\long\def\realizationf{\zeta_{\realization}}%
\global\long\def\periodicf{\zeta_{\mathrm{p}}}%
\global\long\def\realizations#1{#1_{y}}%
\global\long\def\ensembleint#1{\int_{\cell}#1\mathrm{d}\realization}%
\global\long\def\periodic#1{#1_{\mathrm{p}}}%
\global\long\def\pl#1{#1^{+}}%
\global\long\def\mn#1{#1^{-}}%
\global\long\def\scalarG{G}%
\global\long\def\zerosub#1{#1_{0}}%
\global\long\def\parameter{y}%
\global\long\def\samplespace{\Lambda}%
\global\long\def\scalarelas{C}%
\global\long\def\scalarpiezo{B}%
\global\long\def\scalarA{A}%
\global\long\def\scalarS{S}%
\global\long\def\scalarW{W}%
\global\long\def\effectivemass{\tilde{\boldsymbol{\rho}}}%

\global\long\def\constsub#1{#1_{c}}%
\global\long\def\wronskian{V}%
\global\long\def\eigens{\eta}%

\global\long\def\real{\mathbb{R}}%
\global\long\def\RefVol{\Omega_{0}}%
\global\long\def\Refx{\mathbf{X}}%
\global\long\def\Curx{\mathbf{x}}%
\global\long\def\map{\boldsymbol{\chi}}%
\global\long\def\defgrad{\mathbf{F}}%
\global\long\def\defgradT{\mathbf{F}^{\mathrm{T}}}%
\global\long\def\defgradTi{\mathbf{F}^{\mathrm{-T}}}%
\global\long\def\d{\mathrm{d}}%
\global\long\def\RCG{\mathbf{C}}%
\global\long\def\LCG{\mathbf{b}}%
\global\long\def\div#1{\nabla\cdot#1}%
\global\long\def\curl#1{\nabla\times#1}%
\global\long\def\T#1{#1^{\mathrm{T}}}%
\global\long\def\CurStress{\boldsymbol{\sigma}}%
\global\long\def\Refcurl#1{\nabla_{\mathbf{X}}\times#1}%
\global\long\def\s#1{#1^{\star}}%
\global\long\def\m#1{#1^{(m)}}%
\global\long\def\f#1{#1^{(f)}}%
\global\long\def\p#1{#1^{(p)}}%
\global\long\def\Refdiv#1{\nabla_{\mathbf{X}}\cdot#1}%
\global\long\def\phase#1{#1^{\left(p\right)}}%
\global\long\def\jm{j_{m}^{(p)}}%
\global\long\def\xinc{\mathbf{\dot{x}}}%
\global\long\def\Pinc{\mathbf{\dot{P}}}%
\global\long\def\Einc{\mathbf{\dot{E}}}%
\global\long\def\Finc{\mathbf{\dot{F}}}%
\global\long\def\dinc{\mathbf{\check{d}}}%
\global\long\def\einc{\mathbf{\check{e}}}%
\global\long\def\Dinc{\mathbf{\dot{D}}}%
\global\long\def\sinc{\boldsymbol{\Sigma}}%
\global\long\def\fA{\mathbb{\boldsymbol{\mathscr{A}}}}%
\global\long\def\fB{\mathbb{\boldsymbol{\mathscr{B}}}}%
\global\long\def\fC{\mathbb{\boldsymbol{\mathscr{C}}}}%
\global\long\def\plane{\left(x_{1},x_{3}\right)}%
\global\long\def\anti{\dot{x}_{2}}%
\global\long\def\bnd{\partial\Omega}%
\global\long\def\rcg{\mathbf{C}}%
\global\long\def\lcg{\mathbf{b}}%
\global\long\def\rcgcomp#1{C_{#1}}%
\global\long\def\cronck#1{\delta_{#1}}%
\global\long\def\lcgcomp#1{b_{#1}}%
\global\long\def\deformation{\boldsymbol{\chi}}%
\global\long\def\cauchycomp#1{\sigma_{#1}}%
\global\long\def\idgt{\dg^{\mathrm{-T}}}%
\global\long\def\cauchy{\boldsymbol{\sigma}}%
\global\long\def\normal{\mathbf{n}}%
\global\long\def\normall{\mathbf{N}}%
\global\long\def\traction{\mathbf{t}}%
\global\long\def\tractionl{\mathbf{t}_{L}}%
\global\long\def\ed{\mathbf{d}}%
\global\long\def\edcomp#1{d_{#1}}%
\global\long\def\edl{\mathbf{D}}%
\global\long\def\edlcomp#1{D_{#1}}%
\global\long\def\ef{\mathbf{e}}%
\global\long\def\efcomp#1{e_{#1}}%
\global\long\def\efl{\mathbf{E}}%
\global\long\def\freech{q_{e}}%
\global\long\def\surfacech{w_{e}}%
\global\long\def\outer#1{#1^{\star}}%
\global\long\def\perm{\epsilon_{0}}%
\global\long\def\matper{\epsilon}%
\global\long\def\jump#1{\llbracket#1\rrbracket}%
\global\long\def\identity{\mathbf{I}}%
\global\long\def\area{\mathrm{d}a}%
\global\long\def\areal{\mathrm{d}A}%
\global\long\def\refsys{\mathbf{X}}%
\global\long\def\Grad{\nabla_{\refsys}}%
\global\long\def\grad{\nabla}%
\global\long\def\divg{\nabla\cdot}%
\global\long\def\Div{\nabla_{\refsys}}%
\global\long\def\derivative#1#2{\frac{\partial#1}{\partial#2}}%
\global\long\def\aef{\Psi}%
\global\long\def\dltendl{\edl\otimes\edl}%
\global\long\def\tr#1{\mathrm{tr}#1}%
\global\long\def\ii#1{I_{#1}}%
\global\long\def\dh{\hat{D}}%
\global\long\def\inc#1{\dot{#1}}%
\global\long\def\sys{\mathbf{x}}%
\global\long\def\curl{\nabla}%
\global\long\def\Curl{\nabla_{\refsys}}%
\global\long\def\piolaincpush{\boldsymbol{\Sigma}}%
\global\long\def\piolaincpushcomp#1{\Sigma_{#1}}%
\global\long\def\elaspush{\boldsymbol{\mathcal{C}}}%
\global\long\def\elecpush{\boldsymbol{\mathcal{A}}}%
\global\long\def\elaselecpush{\boldsymbol{\mathcal{B}}}%
\global\long\def\disgrad{\mathbf{h}}%
\global\long\def\disgradcomp#1{h_{#1}}%
\global\long\def\trans#1{#1^{\mathrm{\mathsf{T}}}}%
\global\long\def\phase#1{#1^{\left(p\right)}}%
\global\long\def\elecpushcomp#1{\mathcal{A}_{#1}}%
\global\long\def\elaselecpushcomp#1{\mathcal{B}_{#1}}%
\global\long\def\elaspushcomp#1{\mathcal{C}_{#1}}%
\global\long\def\dnh{\aef_{DH}}%
\global\long\def\woo{\varsigma}%
\global\long\def\wif{\Lambda}%
\global\long\def\structurefun{S}%
\global\long\def\dg{\mathbf{F}}%
\global\long\def\dgcomp#1{F_{#1}}%
\global\long\def\piola{\mathbf{P}}%
\global\long\def\refbody{\Omega_{0}}%
\global\long\def\refbnd{\partial\refbody}%
\global\long\def\bnd{\partial\Omega}%
\global\long\def\rcg{\mathbf{C}}%
\global\long\def\lcg{\mathbf{b}}%
\global\long\def\rcgcomp#1{C_{#1}}%
\global\long\def\cronck#1{\delta_{#1}}%
\global\long\def\lcgcomp#1{b_{#1}}%
\global\long\def\deformation{\boldsymbol{\chi}}%
\global\long\def\dgt{\dg^{\mathrm{T}}}%
\global\long\def\idgcomp#1{F_{#1}^{-1}}%
\global\long\def\velocity{\mathbf{v}}%
\global\long\def\accel{\mathbf{a}}%
\global\long\def\vg{\mathbf{l}}%
\global\long\def\idg{\dg^{-1}}%
\global\long\def\cauchycomp#1{\sigma_{#1}}%
\global\long\def\idgt{\dg^{\mathrm{-T}}}%
\global\long\def\cauchy{\boldsymbol{\sigma}}%
\global\long\def\normal{\mathbf{n}}%
\global\long\def\normall{\mathbf{N}}%
\global\long\def\traction{\mathbf{t}}%
\global\long\def\tractionl{\mathbf{t}_{L}}%
\global\long\def\ed{\mathbf{d}}%
\global\long\def\edcomp#1{d_{#1}}%
\global\long\def\edl{\mathbf{D}}%
\global\long\def\edlcomp#1{D_{#1}}%
\global\long\def\ef{\mathbf{e}}%
\global\long\def\efcomp#1{e_{#1}}%
\global\long\def\efl{\mathbf{E}}%
\global\long\def\freech{q_{e}}%
\global\long\def\surfacech{w_{e}}%
\global\long\def\outer#1{#1^{\star}}%
\global\long\def\perm{\epsilon_{0}}%
\global\long\def\matper{\epsilon}%
\global\long\def\jump#1{\llbracket#1\rrbracket}%
\global\long\def\identity{\mathbf{I}}%
\global\long\def\area{\mathrm{d}a}%
\global\long\def\areal{\mathrm{d}A}%
\global\long\def\refsys{\mathbf{X}}%
\global\long\def\Grad{\nabla_{\refsys}}%
\global\long\def\grad{\nabla}%
\global\long\def\divg{\nabla\cdot}%
\global\long\def\Div{\nabla_{\refsys}}%
\global\long\def\derivative#1#2{\frac{\partial#1}{\partial#2}}%
\global\long\def\aef{\Psi}%
\global\long\def\dltendl{\edl\otimes\edl}%
\global\long\def\tr#1{\mathrm{tr}\left(#1\right)}%
\global\long\def\ii#1{I_{#1}}%
\global\long\def\dh{\hat{D}}%
\global\long\def\lage{\mathbf{E}}%
\global\long\def\inc#1{\dot{#1}}%
\global\long\def\sys{\mathbf{x}}%
\global\long\def\curl{\nabla}%
\global\long\def\Curl{\nabla_{\refsys}}%
\global\long\def\piolaincpush{\boldsymbol{\Sigma}}%
\global\long\def\piolaincpushcomp#1{\Sigma_{#1}}%
\global\long\def\edlincpush{\check{\mathbf{d}}}%
\global\long\def\edlincpushcomp#1{\check{d}_{#1}}%
\global\long\def\efincpush{\check{\mathbf{e}}}%
\global\long\def\efincpushcomp#1{\check{e}_{#1}}%
\global\long\def\elaspush{\boldsymbol{\mathcal{C}}}%
\global\long\def\elecpush{\boldsymbol{\mathcal{A}}}%
\global\long\def\elaselecpush{\boldsymbol{\mathcal{B}}}%
\global\long\def\disgrad{\mathbf{h}}%
\global\long\def\disgradcomp#1{h_{#1}}%
\global\long\def\trans#1{#1^{\mathrm{T}}}%
\global\long\def\phase#1{#1^{\left(p\right)}}%
\global\long\def\elecpushcomp#1{\mathcal{A}_{#1}}%
\global\long\def\elaselecpushcomp#1{\mathcal{B}_{#1}}%
\global\long\def\elaspushcomp#1{\mathcal{C}_{#1}}%
\global\long\def\dnh{\aef_{DG}}%
\global\long\def\dnhc{\mu\lambda^{2}}%
\global\long\def\dnhcc{\frac{\mu}{\lambda^{2}}+\frac{1}{\matper}d_{2}^{2}}%
\global\long\def\dnhb{\frac{1}{\matper}d_{2}}%
\global\long\def\afreq{\omega}%
\global\long\def\dispot{\phi}%
\global\long\def\edpot{\varphi}%
\global\long\def\kh{\hat{k}}%
\global\long\def\afreqh{\hat{\afreq}}%
\global\long\def\phasespeed{c}%
\global\long\def\bulkspeed{c_{B}}%
\global\long\def\speedh{\hat{c}}%
\global\long\def\dhth{\dh_{th}}%
\global\long\def\bulkspeedl{\bulkspeed_{\lambda}}%
\global\long\def\khth{\hat{k}_{th}}%
\global\long\def\p#1{#1^{\left(p\right)}}%
\global\long\def\maxinccomp#1{\inc{\outer{\sigma}}_{#1}}%
\global\long\def\maxcomp#1{\outer{\sigma}_{#1}}%
\global\long\def\relper{\matper_{r}}%
\global\long\def\sdh{\hat{d}}%
\global\long\def\iee{\varphi}%
\global\long\def\effectivemu{\tilde{\mu}}%
\global\long\def\fb#1{#1^{\left(a\right)}}%
\global\long\def\mt#1{#1^{\left(b\right)}}%
\global\long\def\phs#1{#1^{\left(p\right)}}%
\global\long\def\thc{h}%
\global\long\def\state{\mathbf{s}}%
\global\long\def\harmonicper{\breve{\matper}}%
\global\long\def\kb{k_{B}}%
\global\long\def\cb{\bar{c}}%
\global\long\def\mb{\bar{\mu}}%
\global\long\def\rb{\bar{\rho}}%
\global\long\def\wavenumber{k}%
\global\long\def\nh{\mathbf{n}}%
\global\long\def\mh{\mathbf{m}}%
\global\long\def\deflect{\inc x_{2}}%
\global\long\def\sdd#1{#1_{2,11}}%
\global\long\def\sdddd#1{#1_{2,1111}}%
\global\long\def\sd#1{#1_{2,1}}%
\global\long\def\sddd#1{#1_{2,111}}%
\global\long\def\xdddd#1{#1_{,\xi\xi\xi\xi}}%
\global\long\def\xdd#1{#1_{,\xi\xi}}%
\global\long\def\xd#1{#1_{,\xi}}%
\global\long\def\xddd#1{#1_{,\xi\xi\xi}}%
\global\long\def\jm{J_{m}}%
\global\long\def\dv{\Delta V}%
\global\long\def\ih{\mathbf{i}_{1}}%
\global\long\def\kh{\mathbf{i}_{3}}%
\global\long\def\jh{\mathbf{i}_{2}}%
\global\long\def\etil{E}%
\global\long\def\genT{\mathsf{Q}}%
\global\long\def\transfer{\mathsf{T}}%
\global\long\def\statevec{\mathbf{s}}%
\global\long\def\coefvec{\mathbf{c}}%
\global\long\def\pressure{p_{0}}%
\global\long\def\ncell#1{#1_{\left(n\right)}}%
\global\long\def\ydisp{\inc x_{2}}%
\global\long\def\ycord{x_{2}}%
\global\long\def\pn#1{\ncell{#1}^{\left(p\right)}}%
\global\long\def\pnm#1{#1_{\left(n\right)m}^{\left(p\right)}}%
\global\long\def\eigen{\boldsymbol{\eta}}%
\global\long\def\xcomp{x_{1}}%
\global\long\def\totalT{\mathsf{T_{\mathrm{tot}}}}%
\global\long\def\rads{\frac{\mathrm{rad}}{\mathrm{s}}}%
\global\long\def\lf{\gamma}%
\global\long\def\tf{T_{m}}%
\global\long\def\eigenim{\beta}%
\global\long\def\bS{\mathsf{S}}%
\global\long\def\Blochwn{{k}_{\mathit{B}}}%
\global\long\def\Blochwnx{k_{B1}}%
\global\long\def\Blochwny{k_{B2}}%
\global\long\def\waven{k}%
\global\long\def\Beff{\tilde{B}}%
\global\long\def\rhoaux{\check{\rho}}%
\global\long\def\nuaux{\check{\nu}}%
\global\long\def\Daux{\check{D}}%
\global\long\def\geff{\tilde{g}}%
\global\long\def\avevarsigma{\overline{\varsigma}}%
\global\long\def\Deff{\tilde{D}}%
\global\long\def\Dxeff{\tilde{D}_{1}}%
\global\long\def\Dyeff{\tilde{D}_{2}}%
\global\long\def\Dnueff{\tilde{D}_{\nu}}%
\global\long\def\Dxyeff{\tilde{D}_{12}}%
\global\long\def\nueff{\tilde{\nu}}%
\global\long\def\rhoeff{\tilde{\rho}}%
\global\long\def\aveR{\overline{R}}%
\global\long\def\aveMx{\overline{M}_{1}}%
\global\long\def\aveMy{\overline{M}_{2}}%
\global\long\def\aveMxy{\overline{M}_{12}}%
\global\long\def\aveM{\overline{M}}%
\global\long\def\avew{\overline{w}}%
\global\long\def\avep{\overline{p}}%
\global\long\def\vectora{\boldsymbol{\psi}}%
\global\long\def\vectorb{\boldsymbol{\varphi}}%
\global\long\def\vectorp{\mathsf{P}}%
\global\long\def\avevectora{\overline{\boldsymbol{\psi}} }%
\global\long\def\avevectorao{\left\langle \boldsymbol{\psi}_{0}\right\rangle }%
\global\long\def\avevectorb{\overline{\boldsymbol{\varphi}} }%
\global\long\def\avevectorp{\overline{\mathsf{P}} }%
\global\long\def\avevectorax{\left\langle \boldsymbol{\psi}{_{,x}}\right\rangle }%
\global\long\def\avevectorat{\left\langle \boldsymbol{\psi}{_{,t}}\right\rangle }%
\global\long\def\avevectorbx{\left\langle \boldsymbol{\varphi}{_{,x}}\right\rangle }%
\global\long\def\DDeff{\tilde{\tilde{D}}}%
\global\long\def\DDxeff{\tilde{\tilde{D}}_{1}}%
\global\long\def\DDyeff{\tilde{\tilde{D}}_{2}}%
\global\long\def\DDnueff{\tilde{\tilde{D}}_{\nu}}%
\global\long\def\nuueff{\tilde{\tilde{\nu}}}%
\global\long\def\DDkeff{\tilde{\tilde{D}}_{k}}%
\global\long\def\DDxyeff{\tilde{\tilde{D}}_{12}}%
\global\long\def\rhooeff{\tilde{\tilde{\rho}}}%
\global\long\def\angRR{\left\langle \left\langle R_{\mathrm{p}}\right\rangle \right\rangle }%
\global\long\def\angMMx{\left\langle M_{1\mathrm{p}}\right\rangle }%
\global\long\def\angMMy{\left\langle M_{2\mathrm{p}}\right\rangle }%
\global\long\def\angMMxy{\left\langle M_{12\mathrm{p}}\right\rangle }%
\global\long\def\angww{\left\langle w_{\mathrm{p}}\right\rangle }%
\global\long\def\angpp{ \left\langle p_{\mathrm{p}}\right\rangle }%
\global\long\def\uF{u_{\mathrm{F}}}%
\global\long\def\wF{w_{\mathrm{F}}}%
\global\long\def\rhoF{\rho_{\mathrm{F}}}%
\global\long\def\BF{B_{\mathrm{F}}}%
\global\long\def\dl{l^{(i)}}%
\global\long\def\bF{\state_{\mathrm{d}}}%
\global\long\def\bA{\state_{\mathrm{f}}}%
\global\long\def\ba{\coefvec}%
\global\long\def\bcero{\mathsf{0}}%
\global\long\def\bH{\mathsf{H}}%
\global\long\def\ltotal{l_{\mathrm{tot}}}%
\global\long\def\bI{\mathsf{I}}%
\global\long\def\qpd{\genT_{\mathrm{d}}^{+}}%
\global\long\def\qmd{\genT_{\mathrm{d}}^{-}}%
\global\long\def\qpf{\genT_{\mathrm{f}}^{+}}%
\global\long\def\qmf{\genT_{\mathrm{f}}^{-}}%
\global\long\def\curcon{\Omega}%
\global\long\def\ld{\mathbf{D}}%
\global\long\def\qpdz{\genT_{\mathrm{d}\left(0\right)}^{+}}%
\global\long\def\qmdz{\genT_{\mathrm{d}\left(0\right)}^{-}}%
\global\long\def\qpfz{\genT_{\mathrm{f}\left(0\right)}^{+}}%
\global\long\def\qmfz{\genT_{\mathrm{f}\left(0\right)}^{-}}%
\global\long\def\qpdl{\genT_{\mathrm{d}\left(M\right)}^{+}}%
\global\long\def\qmdl{\genT_{\mathrm{d}\left(M\right)}^{-}}%
\global\long\def\qpfl{\genT_{\mathrm{f}\left(M\right)}^{+}}%
\global\long\def\qmfl{\genT_{\mathrm{f}\left(M\right)}^{-}}%
\global\long\def\e{\mathop{{\rm \mbox{{\Large e}}}}\nolimits}%
\global\long\def\Tr{\textrm{Tr}}%
\global\long\def\Det{\textrm{Det}}%
\global\long\def\sgn{\textrm{sgn}}%
\global\long\def\pr{^{\prime}}%
\global\long\def\bn#1{\mbox{\boldmath\ensuremath{#1}}}%
\global\long\def\bB{\bn B}%
\global\long\def\bP{\bn P}%
\global\long\def\bY{\bn Y}%
\global\long\def\bV{\bn V}%
\global\long\def\bW{\bn W}%
\global\long\def\bG{\bn G}%
\global\long\def\bM{\bn M}%
\global\long\def\bm{\bn m}%
\global\long\def\bE{\bn E}%
\global\long\def\bK{\bn K}%
\global\long\def\bL{\bn L}%
\global\long\def\bC{\bn C}%
\global\long\def\bT{\bn T}%
\global\long\def\bg{\bn g}%
\global\long\def\bN{\bn N}%
\global\long\def\bX{\bn X}%
\global\long\def\bR{\bn R}%
\global\long\def\bHs{\bn{Hs}}%
\global\long\def\bDelta{\bn{\Delta}}%
\global\long\def\bsigma{\bn{\sigma}}%
\global\long\def\bpsi{\bn{\psi}}%
\global\long\def\bQ{\bn Q}%
\global\long\def\bZ{\bn Z}%
\global\long\def\bU{\bn U}%
\global\long\def\bz{\bn z}%
\global\long\def\bu{\bn u}%
\global\long\def\bk{\bn k}%
\global\long\def\bUpsilon{\bn{\Upsilon}}%
\global\long\def\bEta{\bn{\eta}}%
\global\long\def\bmu{\bn{\mu}}%
\global\long\def\brho{\bn{\rho}}%
\global\long\def\bpartial{\bn{\partial}}%
\global\long\def\Elecf{\mathbf{E}}%
\global\long\def\Edisplacement{\mathbf{D}}%
\global\long\def\Strain{\grad\disp}%
\global\long\def\Stress{\boldsymbol{\sigma}}%
\global\long\def\Piezoelectricmodule{\mathbf{B}}%
\global\long\def\Permittivity{\mathbf{A}}%
\global\long\def\alfacoupled{\check{\alpha}}%
\global\long\def\gammacoupled{\check{\gamma}}%
\global\long\def\bettacoupled{\check{\beta}}%
\global\long\def\rhocoupled{\check{\rho}}%
\global\long\def\Scoupled{\check{S}}%
\global\long\def\Lcoupled{\check{L}}%
\global\long\def\Qcoupled{\check{Q}}%
\global\long\def\elas{\mathbf{C}}%
\global\long\def\momentum{\mathbf{p}}%
\global\long\def\force{\mathbf{f}}%
\global\long\def\disp{\mathbf{u}}%
\global\long\def\potential{\phi}%
\global\long\def\state{\mathsf{S}}%
\global\long\def\position{\mathbf{x}}%
\global\long\def\cmat{\mathsf{C}}%
\global\long\def\rmat{\mathsf{R}}%
\global\long\def\mmat{\mathsf{m}}%
\global\long\def\pmat{\mathsf{p}}%
\global\long\def\gradt{\mathsf{B}}%
\global\long\def\genpotential{\mathsf{w}}%
\global\long\def\vol{\Omega}%
\global\long\def\rve{\Omega^{\mathrm{rve}}}%
\global\long\def\effective#1{\tilde{#1}}%
\global\long\def\ensemble#1{\left\langle #1\right\rangle }%
\global\long\def\average#1{\overline{#1}}%
\global\long\def\rg{\mathbf{W}}%
\global\long\def\po#1{#1^{\left(1\right)}}%
\global\long\def\pt#1{#1^{\left(2\right)}}%
\global\long\def\ptr#1{#1^{\left(3\right)}}%
\global\long\def\length{l}%
\global\long\def\dt#1{\dot{#1}}%
\global\long\def\willis{\mathbf{S}}%
\global\long\def\area{a}%
\global\long\def\fmat{\mathsf{f}}%
\global\long\def\lmat{\mathsf{L}}%
\global\long\def\hmat{\mathsf{h}}%
\global\long\def\bmat{\mathsf{b}}%
\global\long\def\green{\mathsf{G}}%
\global\long\def\adjoint#1{#1^{\dagger}}%
\global\long\def\cc#1{#1^{*}}%
\global\long\def\intvol#1{\int_{\vol}#1\mathrm{d}\sys}%
\global\long\def\boundaryw{\bnd_{\mathsf{w}}}%
\global\long\def\intsurfacepot#1{\int_{\boundaryw}#1\mathrm{d}\area}%
\global\long\def\boundaryt{\bnd_{\mathsf{t}}}%
\global\long\def\boundary{\bnd}%
\global\long\def\intsurfaceh#1{\int_{\boundaryt}#1\mathrm{d}\area}%
\global\long\def\normalop{\mathsf{n}}%
\global\long\def\smat{\mathsf{s}}%
\global\long\def\rene#1{#1^{rene}}%
\global\long\def\fmat{\mathsf{f}}%
\global\long\def\dmat{\mathsf{D}}%
\global\long\def\zm#1#2{\mathsf{0}_{#1\times#2}}%
\global\long\def\nmat{\mathsf{N}}%
\global\long\def\tp#1{#1^{\mathsf{T}}}%
\global\long\def\charge{q}%
\global\long\def\scharge{\omega}%
\global\long\def\dirac{\delta}%
\global\long\def\green{\mathsf{G}}%
\global\long\def\adjoint#1{#1^{\dagger}}%
\global\long\def\cc#1{#1^{*}}%
\global\long\def\boundaryw{\bnd_{\mathsf{w}}}%
\global\long\def\intsurfacepot#1{\int_{\boundaryw}#1\mathrm{d}\area}%
\global\long\def\boundaryt{\bnd_{\mathsf{t}}}%
\global\long\def\intsurfaceh#1{\int_{\boundaryt}#1\mathrm{d}\area}%
\global\long\def\normalop{\mathsf{n}}%
\global\long\def\smat{\mathsf{s}}%
\global\long\def\rene#1{#1^{rene}}%
\global\long\def\fmat{\mathsf{f}}%
\global\long\def\dmat{\mathsf{D}}%
\global\long\def\zm#1#2{\mathsf{0}_{#1\times#2}}%
\global\long\def\nmat{\mathsf{N}}%
\global\long\def\tp#1{#1^{\mathsf{T}}}%
\global\long\def\charge{q}%
\global\long\def\scharge{\omega}%
\global\long\def\dirac{\delta}%
\global\long\def\period{l}%
\global\long\def\scalarf{f}%
\global\long\def\scalarD{D}%
\global\long\def\randomvar{p}%
\global\long\def\cell{\vol_{\mathrm{p}}}%
\global\long\def\realization{y}%
\global\long\def\realizationf{\zeta_{\realization}}%
\global\long\def\periodicf{\zeta_{\mathrm{p}}}%
\global\long\def\realizations#1{#1_{y}}%
\global\long\def\ensembleint#1{\int_{\cell}#1\mathrm{d}\realization}%
\global\long\def\periodic#1{#1_{\mathrm{p}}}%
\global\long\def\pl#1{#1^{+}}%
\global\long\def\mn#1{#1^{-}}%
\global\long\def\scalarG{G}%
\global\long\def\zerosub#1{#1_{0}}%
\global\long\def\parameter{y}%
\global\long\def\samplespace{\Lambda}%
\global\long\def\constsub#1{#1_{c}}%
\global\long\def\wronskian{V}%
\global\long\def\stressindex{\sigma}%
\global\long\def\Edisplaindex{D}%
\global\long\def\dispIndex{u}%
\global\long\def\forceindex{f}%
\global\long\def\Gindex{G}%
\global\long\def\elasindex{C}%
\global\long\def\Piezoelectricindex{B}%
\global\long\def\Permittivityindex{A}%
\global\long\def\eigenindex{\eta}%
\global\long\def\nhindex{n}%
\global\long\def\positionscalar{x}%
\global\long\def\plmn#1{#1^{\pm}}%
\global\long\def\coeffG{V}%
\global\long\def\scalarmomentum{p}%
\global\long\def\Piezoelectricmoda{\Piezoelectricmodule_{1}}%
\global\long\def\Piezoelectricmodb{\Piezoelectricmodule_{2}}%
\global\long\def\willisa{\mathcal{S}_{1}}%
\global\long\def\willisb{\mathcal{S}_{2}}%
\global\long\def\rga{\mathcal{W}_{1}}%
\global\long\def\rgb{\mathcal{W}_{2}}%
\global\long\def\greentensor{\mathbf{G}}%
\global\long\def\elascalar{C}%
\global\long\def\Piezoelectricscalar{B}%
\global\long\def\Permittivityscalar{A}%
\global\long\def\williscalar{S}%
\global\long\def\rgscalar{W}%
\global\long\def\scalarelas{C}%
\global\long\def\scalarpiezo{B}%
\global\long\def\scalarA{A}%
\global\long\def\scalarS{S}%
\global\long\def\scalarW{W}%
\global\long\def\sure{\mathsf{a}}%
\global\long\def\equivc{\check{\scalarelas}}%
\global\long\def\phase#1{#1^{\left(n\right)}}%
\global\long\def\constanta{\alpha}%
\global\long\def\constantb{\beta}%
\global\long\def\statev{\mathsf{s}}%
\global\long\def\coefficients{\varsigma}%
\global\long\def\adjointerm{\vartheta}%
\global\long\def\abreviata{\zeta}%
\global\long\def\abreviatb{\gamma}%
\global\long\def\qphase#1{#1^{\left(p\right)}}%
\global\long\def\qphase#1{#1^{\left(q\right)}}%

\global\long\def\ucoefficient{c}%
\global\long\def\reflCoef{r}%

\global\long\def\TransCoeff{t}%
\global\long\def\transfer{\mathsf{T}}%
\global\long\def\CoeffTransferM{\mathsf{K}}%
\global\long\def\Younmodul{E}%
\global\long\def\metaLenfth{d}%

\global\long\def\frequency{\mathrm{f}}%

\global\long\def\Densitycontrast{\lambda}%
\global\long\def\Econtrast{\alpha}%
\global\long\def\Dielecontrast{\beta}%
\global\long\def\Piezocontrast{\gamma}%

\global\long\def\CurrentDensity{\boldsymbol{\mathbf{J}}}%
\global\long\def\magnefield{\boldsymbol{\mathbf{H}}}%
\global\long\def\magneticflux{\mathcal{\boldsymbol{B}}}%
\global\long\def\velocity{\mathbf{v}}%

\global\long\def\magnefieldscalar{H}%
\global\long\def\CurrentDenscalar{J}%
\global\long\def\velocscalar{v}%

\global\long\def\scalarE{E}%
\global\long\def\scalarmagFlux{\mathcal{B}}%
\global\long\def\permeabtensor{\mathbf{\boldsymbol{\mu}}}%

\global\long\def\Generalrespons{\mathbf{\boldsymbol{\chi}}}%
\global\long\def\GeneraldrivForce{\boldsymbol{\mathbf{Y}}}%
\global\long\def\GeneralHistoryfunction{\mathrm{\boldsymbol{\mathbf{Q}}}}%
\global\long\def\Contsmatrix{\mathbf{\mathsf{M}}}%
\global\long\def\MatrixQ{\mathsf{Q}}%

\global\long\def\PropTerm{\Gamma}%
\global\long\def\intvolx#1{\int_{\vol}#1\mathrm{d}x}%
\global\long\def\dtt#1{\ddot{#1}}%
\global\long\def\ComplexPower{P_{\mathrm{c}}}%
\global\long\def\AveragePowerLoss{\ComplexPower'}%
\global\long\def\CausalWillis{\psi}%
\global\long\def\CausalEM{\chi}%
\global\long\def\recip{\Delta P}%
\global\long\def\CompVar{\mathit{z}}%
\global\long\def\Wavevector{\boldsymbol{\kappa}}%
\global\long\def\FTrans#1{\check{#1}}%
\global\long\def\Wavenumber{\kappa}%
\global\long\def\KernelL{\effective{\lmat}}%
\global\long\def\gmat{\mathsf{g}}%
$ $

\global\long\def\ensembleintvect#1{\int_{\cell}#1\mathrm{d}\mathbf{y}}%
\global\long\def\Blochwvect{\mathbf{k}_{\mathit{B}}}%
\global\long\def\yposition{\mathbf{y}}%
\global\long\def\Displgrad{\boldsymbol{\varepsilon}}%
\global\long\def\PDF{P}%
\global\long\def\genpot{\mathsf{w}}%
\global\long\def\dist{l}%
\global\long\def\load#1{#1^{\left(\dist\right)}}%
\global\long\def\bterm{\mathsf{B.T.}}%
\global\long\def\genkcomp{K}%
\global\long\def\genb{b}%
\global\long\def\gend{\Lambda}%
\global\long\def\gendiscomp{w}%
\global\long\def\inner#1#2{\big\langle#1,#2\big\rangle_{\vol}}%
\global\long\def\constop{\mathcal{M}}%
\global\long\def\genpotb{\mathsf{v}}%
\global\long\def\conj#1{#1^{*}}%
\global\long\def\dummyx{\boldsymbol{\chi}}%
\global\long\def\intvoldummy#1{\int_{\vol}#1\mathrm{d}\dummyx}%
\global\long\def\adjointforce{m}%
\global\long\def\scalark{\kappa}%
\global\long\def\oz{\omega_{0}}%
\global\long\def\Funtion{\frac{\Delta\FTrans{\lmat}}{\Delta\omega}}%
\global\long\def\aop{\mathcal{A}}%
\global\long\def\bop{\mathcal{B}}%
\global\long\def\cop{\mathcal{C}}%
\global\long\def\rop{\mathcal{R}}%
\global\long\def\wop{\mathcal{W}}%
\global\long\def\sop{\mathcal{S}}%
\global\long\def\lop{\mathcal{L}}%
\global\long\def\intvolprime#1{\int_{\vol}#1\mathrm{d}\dummyx}%
\global\long\def\modwop{\widehat{\mathcal{W}}}%
\global\long\def\modsop{\widehat{\mathcal{S}}}%
\global\long\def\dummyb{\mathbf{X}}%
\global\long\def\bte{\adjoint{\effective{\Piezoelectricmodule}}}%
\global\long\def\modw{\widehat{\rg}}%
\global\long\def\mods{\widehat{\willis}}%
\global\long\def\modwc{\widehat{W}}%
\global\long\def\modsc{\widehat{S}}%
\global\long\def\Curlp{\nabla_{\position'}}%
\global\long\def\Curlsp{\nabla_{\position}}%
\global\long\def\Hadj#1{#1^{\mathsf{T}*}}%
\global\long\def\Ginv{\mathsf{V}}%
\global\long\def\CausalWillisT{\mathsf{\mods}}%
\global\long\def\CausalEMT{\modw}%
\global\long\def\hpart#1{#1^{H}}%
\global\long\def\nhpart#1{#1^{SH}}%
\global\long\def\qnhpart#1{#1^{QSH}}%
\global\long\def\qhpart#1{#1^{QH}}%
\def\Xint#1{\mathchoice
{\XXint\displaystyle\textstyle{#1}}%
 {\XXint\textstyle\scriptstyle{#1}}%
 {\XXint\scriptstyle\scriptscriptstyle{#1}}%
 {\XXint\scriptscriptstyle\scriptscriptstyle{#1}}%
 \!\int} \def\XXint#1#2#3{{\setbox0=\hbox{$#1{#2#3}{\int}$}
   \vcenter{\hbox{$#2#3$}}\kern-.5\wd0}}
\def\ddashint{\Xint=}
\def\dashint{\Xint-}

\begin{abstract}
Metamaterials whose momentum is constitutively coupled with their
strain show promise in wave manipulation for engineering purposes,
and are called Willis materials. They were discovered using an effective
medium theory which shows that their response is  non-local in space
and time. Recently, we generalized this theory to account for piezoelectricity,
and demonstrated that the effective momentum can depend constitutively
on the electric field, thereby enlarging the design space for metamaterials.
 Here, we develop the mathematical restrictions on the effective
properties of such generalized Willis materials, owing to passivity,
reciprocity and causality. Establishing these restrictions is of fundamental
significance, as they test the validity of theoretical and experimental
results---and applicational importance, since they provide elementary
bounds for the maximal response that potential devices may achieve.
\end{abstract}
\title{Fundamental principles for generalized Willis metamaterials}
\author{René Pernas-Salomón and Gal Shmuel}
\affiliation{Faculty of Mechanical Engineering, Technion--Israel Institute of
Technology, Haifa 32000, Israel}
\maketitle

\section{Introduction}

The response of artificial composites with
specially designed microstructure can fundamentally differ from the
response of their constituents. Such composites are termed \emph{meta}materials,
and their features span various solid facets, including electromagnetic
and mechanical properties \cite{Simovski2009bc,Kriegler2010IEEE,Kadic2013RPP,Wegener2013,bertoldi2017flexible,Kadi2019nrp}.

A prominent thrust in metamaterial design is wave control \cite{Hao2007prl,gonella15,Cummer2016yu,mo2019cnoidal},
where some of the achievements thus far are  wave suppressors, cloaking,
negative refraction, and superlensing \cite{pendry00,Schurig2006-Science,craster2012acoustic,PARNELL2013WM,bigoni2013prb,Srivastava2016jmps,ge2017,Oh2017prapplied,getz2017b,Hou2018PRApplied,bortot2018b,MORINI2019jmps,Lustig2019,Psiachos2019prb}.
These phenomena are often manifestations of anomalous effective properties,
such as negative refractive index and negative mass \cite{shelby2001experimental,liu2007PRL,cheng2008one,lee2010composite,Liu2011APL},
which are analytically determined  using homogenization (effective
medium) theories \cite{milton2002theory,Craster2010PRSA,Torrent2011njp,Srivastava2012prsa,torrent2014prb,Antonakakis2013PRSA,Antonakakis2014JMPS,srivastava2015elastic,Caleap2015NJP,AMIRKHIZI2017,pernassalomon2018jmps,Sridhar2018JMPS}.
Notably, Willis has developed an elastodynamic homogenization theory
which predicts that the momentum and stress can be constitutively
coupled to the strain and the velocity, respectively, by the now termed
Willis couplings \cite{Willis1980jmps,Willis1981WM,Willis1985IJSS,willis1997book,Willis2011PRSA,WILLIS2012MOM,WILLIS2012MRC}.
These effective properties constitute additional degrees of freedom
to manipulate waves, as was demonstrated, \emph{e.g.}, to experimentally
realize asymmetric reflection  and scattering-free refraction \cite{Koo2016nc,Merkel2018prb,Liu2019prx}.

\begin{figure}[t!]
\centering
\includegraphics[width=0.7\textwidth]{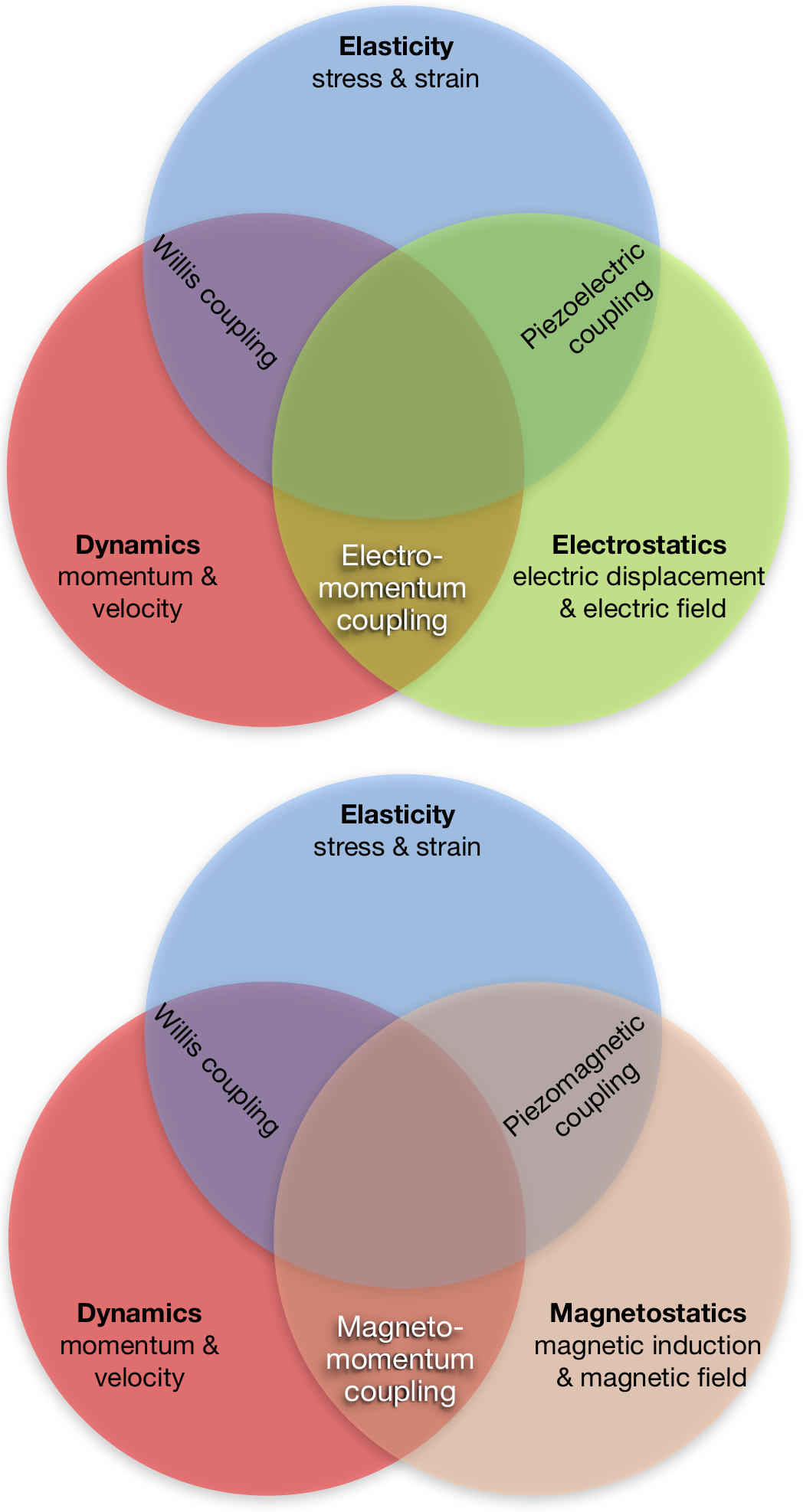} \caption{\label{fig:coupling}Schematics of   the   cross-couplings reported by \citet{PernasSalomon2019JMPS}, in composites whose elasticity is intrinsically coupled with other physics, such as piezoelectric and piezomagnetic materials.
}
\end{figure}

Recently, \citet{PernasSalomon2019JMPS} have generalized the homogenization theory
of Willis to account for constituents that linearly deform in response
to non-mechanical fields, such as piezomagnetic- and piezoelectric
materials \cite{Mason1950fk,Regaz2015ieee}. The main observation
that the generalized theory delivers is the emergence of additional
couplings of Willis type between the momentum and the velocity to
the non-mechanical fields, as illustrated in Fig.$\ $\ref{fig:coupling}.
Accordingly, the momentum of piezoelectric (resp.\ piezomagnetic) composites is coupled with
the electric (resp.\ magnetic) field, while the velocity is coupled with the electric
displacement field (resp.\ magnetic induction). We refer to metamaterials that exhibit these couplings as generalized Willis materials. The additional couplings not only enlarge the design space of metamaterial properties, but also reflect a novel mechanism
to actively manipulate waves via non-mechanical stimuli.

Like all constitutive relations, those that describe Willis materials---standard and generalized---should respect basic physical principles.  \citet{Srivastava2015prsa} and
\citet{Muhlestein20160Prsa2} derived the mathematical restrictions that follow from reciprocity, passivity and causality principles on standard Willis materials in the long-wavelength limit. Here, we continue their works by developing the restrictions that follow from these principles for \emph{generalized }Willis materials.

Accordingly, in the development of the restrictions we account for the coupling between the electric and mechanical governing equations, as well as the additional material properties in the constitutive relations.  Furthermore, the analysis we carry out for reciprocity goes beyond the long-wavelength limit, hence provides new results also for standard Willis materials, additionally to the long-wavelength results in Refs.\ \cite{Srivastava2015prsa,Muhlestein20160Prsa2}.

In the sequel, we show that the obtained mathematical restrictions elucidate the physical nature of such generalized couplings.  These restrictions are also of applicational importance, as pointed out in Refs.\  \cite{Alu2011-PhysRevB,Srivastava2015prsa,Muhlestein20160Prsa2}, since they provide means of testing the
admissibility of experimental data,  and quantifying the maximal response that potential devices
may achieve. For example, \citet{quan2018prl} sought acoustic scatterers with maximum Willis coupling as follows from passivity and reciprocity, and then employed their optimal structure to design metasurfaces for sound steering.

The paper is structured as follows. In the rest of this Sec., we recall
relevant developments in Willis equations, discuss their uniqueness,
and summarize our results before presenting derivations. Sec.$\ $\ref{sec:constitutive-relations}
revisits the theory that lead to the generalized Willis equations,
and introduces a modified formulation, which is motivated by the analysis
in Refs.$\ $\cite{Milton2007njp,nassar2015willis,Muhlestein20160Prsa2}
for the elastic case. Secs.$\ $\ref{sec:Reciprocity-Passivity},
\ref{sec:Reciprocity} and \ref{sec:Causality} develop the restrictions that passivity,
reciprocity and causality pose on the effective relations, respectively. Final comments
conclude this paper in Sec.$\ $\ref{sec:Closure}.

\subsection{\label{subsec:Literature-review}Relevant developments in Willis
equations }

Since this work is closely related to Willis equations, as it provides
physical restrictions on their generalization, a more elaborated review
of their relevant developments is in order. The topics discussed next
do not constitute a complete review of the works in the field, and
aspects such as weighted averages \cite{Milton2007njp,Willis2009},
connections with asymptotic homogenization \cite{NASSAR2016prsa,Meng2018prsa}
etc. are not addressed here.

Willis has started to develop his formulation using a variational
approach that extends the concepts of ensemble averaging and comparison
media  from elastostatics \cite{Willis1981WM,Willis1985IJSS,willatzen2014,Willis1980jmps}.
His effective relations exhibit two notable features, in addition
to the emergence of the cross-couplings mentioned earlier. First,
they are non-local in space---as known from elastostatics---\emph{and
in time,} even if the response of the original composite was history-independent.
(The non-local nature renders the effective relations non-unique,
an issue that is discussed later.) Second, the kernel that describes
the effective mass density is a second-order tensor.

More recently, Willis developed a formulation that does not rely on
a comparison medium, but rather on the Green function of the studied
composite \cite{Willis2011PRSA}. Importantly, he resolved the lack
of uniqueness in the effective properties, which occurs since the
effective strain and velocity are derived from the same potential
(displacement) field. This was carried by adapting the approach of
\citet{FIETZ2010pysicaaB}, which introduces an additional driving
source using an eigenstrain, thereby forcing the effective relative
strain and velocity to be independent. While it is questionable if
such eigenstrains can be experimentally prescribed \cite{Milton2020IV},
their mathematical inclusion has the benefit of providing a unique
effective properties out of an equivalent class that exists when the
eigenstrain vanishes. Source-driven homogenization has been adopted
later also in Refs.$\ $\cite{alu2011prb,Norrisrspa2011PRSA,Sieck2017prb}.
Having listed the main developments in Willis theory, we can now point
out the common components with our theory for media that deform by
non-mechanical stimuli: our theory also relies on ensemble averaging,
incorporates eigenstrains as additional driving source, and delivers
unique effective properties based on the Green function of the original
composite.

The recent interest in metamaterials \footnote{Metamaterials actually have a long history before they earned this
name, see the enlightening summary by \citet{Milton2020II}. } has disseminated to Willis effective relations, resulting with a
bulk of papers that present experimental validation in the long-wavelength limit, and analyze their
structure \cite{nemat2011homogenization,NematNasser2011jmps,Shuvalov2011prsa,nassar2015willis,XIANG2016JMPS,Muhlestein2017nc,NASSAR2017jmps,quan2018prl,su2018prb,Melnikov2019nc,zhai2019arxiv,Lau2019,Liu2019prx,Chen2020nc}.
We list next some of the insights that are relevant to this paper.
\citet{milton06cloaking} identified the similarity between Willis
couplings and bianisotropy in electromagnetics, see also Refs.$\ $\cite{Milton07,quan2018prl,Muhlestein20160Prsa2,Sieck2017prb,su2018prb,Milton2020II}.
\citet{Sieck2017prb} provided a perceptive analysis on the source
of the cross-couplings in periodic media, concluding that the their
non-local part originates from multiple scattering and phase change
at the mesoscale, while their local part originates from asymmetry
in the unit cell. Similarly, \citet{Pernas-Salomon2019} pointed
out the analogy with the broken inversion symmetry in piezoelectric
materials at the atomic scale, which leads to microscopic electroelastic
coupling.

Spatially local couplings were proposed by \citet{milton06cloaking}.
As pointed out in Ref.$\ $\cite{Milton2020II}, the corresponding
equations are the limiting case of the non-local equations, referred
to as the Milton-Briane-Willis equations. According to \citet{Milton2020IV}, the local form is more physical, owing to the difficulty in experimentally measuring unique non-local properties that include the cross-couplings. A local model was developed
by \citet{Milton2007njp}, whose stress depends on the acceleration
rather than the velocity. Simpler spatially local models that report
acceleration-dependent stress were given later in Refs.$\ $\cite{nassar2015willis,Muhlestein20160Prsa2,Milton2016ExtendingTT,Milton2020II}.
These works suggest that the non-local nature of the operator conceals \footnote{The ambiguity associated with the non-local operator is explained
in Sec.$\ $\ref{sec:constitutive-relations}.} a more physical constitutive description---one which employs the
strain rate and acceleration as additional input functions. Here,
we adapt and examine this suggestion to our settings, by introducing
and analyzing a description that additionally includes the time derivative
of the electric field as an input function, and find arguments that
support the use of the alternative formulation.

\subsection{Summary of our results}

As discussed above, in Ref.\ \cite{PernasSalomon2019JMPS} we developed a dynamic homogenization theory for piezoelectric and piezomagnetic composites, which delivers non-local effective relations between suitably defined macroscopic fields.
We formally showed that additional
couplings emerge in the effective relations between the macroscopic momentum and velocity to
the non-mechanical fields. In the sequel, we develop the mathematical restrictions that the effective relations must satisfy in order to respect three principles.

The first principle we analyze is passivity, which at the basic level
means that the material does not generate energy. Formally, we require
that the power supplied  by external agents is always greater or
equal to the rate of change of the energy stored by the material. This principle delivers inequalities for the skew-Hermitian and Hermitian
parts of the Fourier transforms of the effective properties, as summarized
in Tab.$\ $\ref{tab:Tab-restrictions}. If the material exhibits
major symmetries, then these inequalities apply to the imaginary and
real parts of the  transforms. If the material is passive and lossless,
we find that the direct couplings---and combinations of cross-couplings---must
be either Hermitian or skew-Hermitian.

The second principle we employ is reciprocity, which refers to an
equality between the power done by conjugate fields of different problems.
In the long-wavelength limit, it implies major symmetries for direct
couplings, and transpose relations between conjugate cross-coupling terms. This result includes the symmetries reported in Ref.\ \cite{Muhlestein20160Prsa2} for the elastic properties in local Willis materials (i.e., Milton-Briane-Willis materials). Beyond
the long-wavelength, we find that reciprocity requires the non-local operator to be self-adjoint
with respect to the spatial variables. Technically, this translates
to an interchange in the functional dependency in these variables,
in addition to the transposition relations among the couplings (see
Tab.$\ $\ref{tab:Tab-restrictions}).  From this analysis we deduce
that the formulation that does not use the time derivative of the
velocity, strain and electric field is unphysical since it corresponds
to imaginary properties in the time domain. By contrast, the modified
formulation that is based on these rates leads to real properties
in the time domain.

The last principle we employ is causality, which means that an effect
(\emph{e.g.}, momentum) cannot precede its cause (e.g., electric field).
This principle provides a connection between the real and imaginary
parts of the (time) transforms of the couplings. The process we employ
is standard and straightforward, and uses the Plemelj formulas to
obtain relations of the Kramers-Krönig type for the generalized effective
properties \cite{Kronig1926xy,land&lifs60book,Nistad2008PRB,Silveirinha2011PhysRevB}.
We clarify that our study of causality is restricted to the spatially
local equations, and note that the corresponding analysis supports
the claim that the alternative formulation should be favored.

\section{Dynamic homogenization of piezoelectric composites\label{sec:constitutive-relations}}

\begin{figure}[t!]
\centering
\includegraphics[width=.8\textwidth]{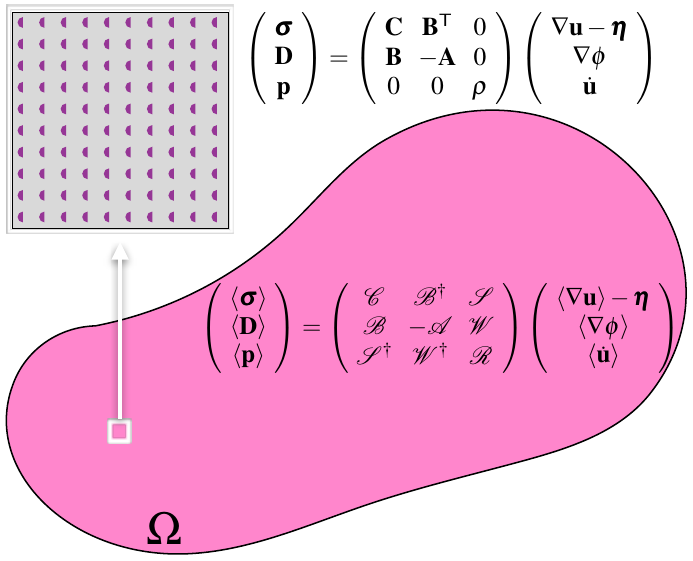} \caption{\label{fig:effective}The body $\Omega$ is composed of different piezoelectric materials whose constitutive response is given by Eq.~\eqref{eq:ConstiRel-1}, as illustrated at the top of the sketch. Effectively, the response of the body is non-local with additional cross-couplings, as given by Eq.~\eqref{eq:EffConstRel}.}
\end{figure}

  We consider a composite occupying the volume $\vol$ made of piezoelectric
phases, driven by time-dependent body force density $\force$, inelastic
strain $\eigen$, and free charge density $\charge$. These sources
generate in the composite stress $\Stress$, electric displacement
$\Edisplacement$ and momentum density $\momentum$, which satisfy
the balance equations
\begin{equation}
\begin{aligned}\div{\Stress}+\force-\dt{\momentum}=\mathbf{0},\end{aligned}
\label{NewtonLaw}
\end{equation}
and
\begin{equation}
\grad\cdot\Edisplacement=\charge,\label{GaussLaw}
\end{equation}
where the superposed dot denotes a time derivative. At each material
point $\sys$, these fields are related to the displacement gradient
$\grad\disp$, velocity $\dt{\disp}$, and electric potential gradient \footnote{In this way, the Faraday equation for the electric field in the quasi-electrostatic
approximation is identically satisfied. This approximation means that
the difference between time scales of the mechanical and electromagnetic
effects allows to neglect time derivatives and magnetic fields in
Maxwell equations.} $\grad\potential$ through the constitutive equations of piezoelectricity
\citep{auld1973acoustic}, namely \footnote{Eq.$\ $\ref{eq:ConstiRel-1} is to be interpreted using the proper
products the matrix elements and the column vector, \emph{e.g.}, $\elas:\eigen$
is a double contraction, which reads in components $C_{ijkl}\eta_{kl}$,
$\tp{\Piezoelectricmodule}\cdot\grad\potential$ is a single contraction,
etc.},
\begin{equation}
\left(\begin{array}{c}
\Stress\\
\Edisplacement\\
\momentum
\end{array}\right)=\left(\begin{array}{ccc}
\elas & \tp{\Piezoelectricmodule} & 0\\
\Piezoelectricmodule & -\Permittivity & 0\\
0 & 0 & \rho
\end{array}\right)\left(\begin{array}{c}
\grad\disp-\eigen\\
\grad\potential\\
\dt{\disp}
\end{array}\right),\label{eq:ConstiRel-1}
\end{equation}
where $\rho$, $\Permittivity$ and $\Piezoelectricmodule$, and $\elas$,
are the spatially varying \footnote{Dependency in $\sys$ is omitted from Eq.$\ $\ref{eq:ConstiRel-1} for brevity.} local mass density, dielectric, piezoelectric and elasticity tensor
fields, respectively \footnote{This is a local material, since the resultant fields at a certain point and time depend only on the variables in that point and time.  If the response of the constituents depend on the history, e.g.,
they are viscoelastic, then the material is temporally non-local, and the constitutive relations become convolutions with
respect to time.}. In co-ordinates, these tensors satisfy
\begin{align}
\scalarA_{ij}=\scalarA_{ji},\scalarpiezo_{ijk}=\scalarpiezo_{jik},\tp{\scalarpiezo}_{ijk}=\scalarpiezo_{kij},\nonumber \\
\scalarelas_{ijkl}=\scalarelas_{jikl}=\scalarelas_{jilk}=\scalarelas_{klij},\sigma_{ij}=\sigma_{ji}.\label{eq:symmetries}
\end{align}
 \citet{PernasSalomon2019JMPS} have proposed an effective description
with constitutive equations for the composite by extending the approach
of \citet{Willis2011PRSA}. This was carried out by treating the composite
as random, such that its properties are not only functions of $\sys$,
but also of the particular specimen that belongs to some sample space
$\samplespace$. The expectation value of any property, say $\rho$,
is given by the ensemble average
\begin{equation}
\ensemble{\rho}\left(\sys\right)=\int_{\samplespace}\rho\left(\sys,y\right)\PDF\left(y\right)\mathrm{d}y,\label{eq:ensemble averagre}
\end{equation}
where the parameter $y$ is used to label the specimens, and $\PDF$
is the probability measure function over $\samplespace$. The governing
equations of our effective description are given by the following
ensemble averages of Eqs.~\eqref{NewtonLaw} and \eqref{GaussLaw}
\begin{equation}
\begin{aligned}\grad\cdot\ensemble{\Stress}+\force-\dt{\ensemble{\momentum}}=\mathbf{0},\end{aligned}
\quad\grad\cdot\ensemble{\Edisplacement}=\charge,\label{ensembledeqs}
\end{equation}
in which $\ensemble{\Stress},\ensemble{\Edisplacement}$ and $\ensemble{\momentum}$
are the effective fields \footnote{Note that $\left<\force\right>=\force$,  $\left<q\right>=\charge$ and $\left<\eta\right>=\eta$ since $\force$, $\charge$ and $\eta$ are taken as sure.}.
Based on the Green (tensor) function of
the problem, \citet{PernasSalomon2019JMPS} obtained constitutive
equations for the effective fields in the form \footnote{Note that in Eq.$\ $\ref{eq:ConstiRel-1} we amend our notation for the coupling between $\left<\mathbf{\sigma}\right>$ and $\left<\nabla\potential\right>$ in Ref.~\citet{PernasSalomon2019JMPS} , by changing the superscript $\mathsf{T}$ to $\dagger$.}
\begin{equation}
\left(\begin{array}{c}
\ensemble{\Stress}\\
\ensemble{\Edisplacement}\\
\ensemble{\momentum}
\end{array}\right)=\left(\begin{array}{ccc}
\cop & \adjoint{\bop} & \sop\\
\bop & -\aop & \wop\\
\adjoint{\sop} & \adjoint{\wop} & \rop
\end{array}\right)\left(\begin{array}{c}
\ensemble{\grad\disp}-\eigen\\
\ensemble{\grad\potential}\\
\ensemble{\dt{\disp}}
\end{array}\right),\label{eq:EffConstRel}
\end{equation}
where the matrix elements are now \emph{non-local} \emph{operators}
in \emph{time} \emph{and} \emph{space}.
(At this point, we do not
endow the couplings with superscript $\dagger$ the meaning that this
symbol usually designates, and postpone it to Sec.$\ $\ref{sec:Reciprocity}.)
 We denote the column vectors in the left- and right-hand sides
of Eq.$\ $\eqref{eq:EffConstRel} by $\ensemble{\hmat}$ and $\ensemble{\gmat}$,
and put the latter statement into formal footing, namely,
\begin{equation}
\ensemble{\hmat}\left(\position,t\right)=\lop\left(\ensemble{\gmat}\right)=\int_{-\infty}^{t}\intvoldummy{\effective{\lmat}\left(\position,\dummyx,t-T\right)\ensemble{\gmat}\left(\dummyx,T\right)\mathrm{d}T},\label{eq:Time-convolution}
\end{equation}
where $\lop$ denotes the non-local effective constitutive operator
and $\effective{\lmat}$ is its kernel. In the sequel, we will denote
by $\effective{\elas}$ the kernel of $\cop$, by $\effective{\willis}$
the kernel of $\sop$, by $\effectivemass$ the kernel of $\rop$,
etc. The effective operator exhibits three notable features, in addition
to its spatio-temporal non-local nature. First, it couples $\ensemble{\Stress}$
with $\ensemble{\dt{\disp}}$, and $\ensemble{\momentum}$ with $\ensemble{\grad\disp}$,
through the so-called Willis couplings $\sop$ and $\adjoint{\sop}$.
Second, (kernel of) the effective mass density $\effectivemass$ is
a second-order tensor. As mentioned, these two features---which are
absent from the local constitutive equations and hence represent \emph{meta}materials---were
discovered by \citet{Willis1980jmps,Willis1981WM,Willis1985IJSS}
in his studies of purely elastic composites. The third distinctive
feature reported by \citet{PernasSalomon2019JMPS} is the coupling
$\wop$ between $\ensemble{\Edisplacement}$ and $\ensemble{\dt{\disp}}$,
and the coupling $\adjoint{\wop}$ between $\ensemble{\momentum}$
and $\ensemble{\grad\potential}$, which we term the electro-momentum
coupling. The transition to this effective description is schematically illustrated in Fig.~\ref{fig:effective}.
The kernel of $\lop$ is endowed with the minor symmetries
\begin{eqnarray}
 & \effective{\elascalar}_{ijkl}=\effective{\elascalar}_{jikl},\effective{\elascalar}_{ijkl}=\effective{\elascalar}_{ijlk},\effective{\Piezoelectricscalar}_{ijk}=\effective{\Piezoelectricscalar}_{ikj},\adjoint{\effective{\Piezoelectricscalar}_{ijk}}=\adjoint{\effective{\Piezoelectricscalar}_{jik}}\nonumber \\
 & \effective{\williscalar}_{ijk}=\effective{\williscalar}_{jik},\adjoint{\effective{\williscalar}_{ikl}}=\adjoint{\effective{\williscalar}_{ilk}},\label{eq:minorSym}
\end{eqnarray}
as they translate from the microscopic to the effective description,
owing to the balance of angular momentum and independence from the
anti-symmetric part of $\grad\disp$. The major symmetries of the
constitutive tensors in Eq.$\ $\eqref{eq:symmetries} induce additional
symmetries between the effective tensors (and justify the superscript
$\dagger$ mentioned above), to be discussed in Sec.$\ $\ref{sec:Reciprocity}
and the Appendix. 

 When the composite is statistically homogeneous, the constitutive
operator becomes translation invariant, \emph{i.e.}, it depends only
on the difference $\position-\dummyx$; accordingly, Eq.$\ $\eqref{eq:Time-convolution}
has the form of a convolution not only in time, but also in space.
Therefore, the Fourier transform with respect to both time and space
yields constitutive relations in the form of simple products between
the transforms of $\effective{\lmat}$ and $\ensemble{\gmat}$. It
follows that such an infinite medium admits plane waves in the form
of (the real part of) $\ensemble{\disp}=\mathbf{U}e^{i\left(\Wavevector_{0}\cdot\position-\omega_{0}t\right)}$
and $\ensemble{\potential}=\Phi e^{i(\Wavevector_{0}\cdot\position-\omega_{0}t)}$,
for which the non-local constitutive equations are the simple products
\begin{equation}
\ensemble{\hmat}\left(\position,t\right)=\FTrans{\lmat}\left(-\Wavevector_{0},\omega_{0}\right)\ensemble{\gmat}\left(\position,t\right)\label{eq:simplification}
\end{equation}
in the $\left(\position,t\right)$ space. (Again, the real part of
the equation should be taken.) We emphasize that $\FTrans{\lmat}\left(-\Wavevector_{0},\omega_{0}\right)$
is the space-time Fourier transform of $\effective{\lmat}$ according
to the convention
\begin{equation}
\FTrans{\lmat}\left(\Wavevector,\omega\right)=\int_{\vol}\mathrm{d\position}\int_{\mathbb{R}}\mathrm{d}t\,\effective{\lmat}\left(\position,t\right)e^{i\left(\Wavevector\cdot\position+\omega t\right)},\label{eq:space time Fourier transform definition}
\end{equation}
 evaluated at $\left(-\Wavevector_{0},\omega_{0}\right)$. In order
not to introduce more notations to the already large set used here,
we will also use $\check{\left(\circ\right)}$ for transforms that
are applied only with respect to one of the two variables (time or
space).

The objective of this work is to determine the mathematical restrictions
imposed on relations \eqref{eq:EffConstRel}---and specifically on
the electro-momentum coupling---by the physical principles of reciprocity,
passivity, and causality. In addition to form \eqref{eq:EffConstRel},
we will also analyze the form
\begin{align}
\left(\begin{array}{c}
\ensemble{\Stress}\\
\ensemble{\Edisplacement}\\
\ensemble{\momentum}
\end{array}\right) & =\left(\begin{array}{ccc}
\cop & \adjoint{\bop} & 0\\
\bop & -\aop & 0\\
0 & 0 & \rop
\end{array}\right)\left(\begin{array}{c}
\ensemble{\grad\disp}-\eigen\\
\ensemble{\grad\potential}\\
\ensemble{\dt{\disp}}
\end{array}\right)\nonumber \\
 & +\left(\begin{array}{ccc}
0 & 0 & \modsop\\
0 & 0 & \modwop\\
\adjoint{\modsop} & \adjoint{\modwop} & 0
\end{array}\right)\left(\begin{array}{c}
\ensemble{\grad\dot{\disp}}-\dot{\eigen}\\
\ensemble{\grad\dot{\potential}}\\
\ensemble{\ddot{\disp}}
\end{array}\right),\label{eq:EffConstRel-1}
\end{align}
where the kernel of the time Fourier transform of $\modsop$ is $-\FTrans{\willis}/i\omega$,
the kernel of the transform of $\modwop$ is $-\FTrans{\rg}/i\omega$,
etc. The motivation for this form was mentioned in Sec.$\ $\ref{subsec:Literature-review},
and elaborated next. To this end, it is useful to note that the derivations
that  led \citet{Willis2011PRSA} and \citet{PernasSalomon2019JMPS}
to their non-local operators were carried out after applying the Fourier
transform with respect to time \footnote{More precisely, the Laplace transform with the variable $s$ was used,
which is connected to the Fourier transform via $s=-i\omega$. }, where in the frequency domain the cross-coupling terms are products
that include the term $-i\omega$.   An ambiguity emerges when transforming
back to the time domain: should $-i\omega$ be identified with the
kernel or with the time derivative of $\ensemble{\gmat}$? The former
leads to relations (\ref{eq:EffConstRel}), and the latter to relations
(\ref{eq:EffConstRel-1}). The forthcoming analysis supports form
(\ref{eq:EffConstRel-1}), in agreement with Refs.$\ $\cite{Sieck2017prb,Muhlestein20160Prsa2}.

Before we proceed, we note that a similar ambiguity exists when the
transform is applied with respect to the spatial translation \cite{Sieck2017prb}.
In this case, spatial derivatives turn to products with $i\Wavevector$,
and the inverse transform has the same problem as with the inversion
of products of $i\omega$.  We can now highlight the motivation for
introducing $\eigen$: since it is not derived from a potential, there
is no way to ``pull outside'' the gradient operator in order to
obtain the effective displacement field, and then mistake the effective
velocity for the effective strain by multiplying and dividing by $i\omega$
\cite{nassar2015willis}. Evidently, such operations lead to different
sets of effective properties, and particularly a set without Willis
couplings \footnote{The problem goes beyond ambiguity between velocity and strain, since
it also follows that even in non-local elastostatics the compatibility
of the effective strain with the effective displacement field implies
that there are infinitely many kernels that equivalently relate the
stress and the strain \cite{WILLIS2012MRC}.}. Since clearly the velocity or strain cannot be derived from the
electric potential, there is no need in introducing an ``eigen electric
field'' in our theory \footnote{This is only partially correct, since in view of the previous footnote,
without such an eigen electric field there are infinitely many kernels
that equivalently relate the electric displacement field and electric
field in non-local electrostatics \cite{Milton2020IV}. }. Owing to the ambiguity associated with the non--local operator
and the difficulty to measure the non--local cross-coupling, \citet{Milton2020II,Milton2020IV}
recently advocated either the use of the local cross-coupling, or
the use of a non-local operator that relates the displacements to
the applied force. While we do not pursue this notion here, we note that the available experimental evidence we are aware of for such cross-couplings were obtained when non-local interactions are negligible, e.g., Refs.\ \cite{Muhlestein2017nc,Melnikov2019nc}, thereby support the advocacy of Milton for the local equations.

\section{Passivity \label{sec:Reciprocity-Passivity}}

The term passivity has different uses in the literature. Here, it
is interpreted as in Ref.~\cite{Fettweis2011yu}, namely, a system
is passive if there exists a positive-definite stored energy function
for it, determined uniquely by its state variables, such that the
power supplied to the system by external agents is always greater
or equal to the rate of change of its stored energy. This requirement,
in turn, poses restrictions on the constitutive parameters \cite{banerjee2011introduction,Srivastava2015prsa}.
The implications of passivity were employed in Refs.~\cite{Muhlestein20160Prsa2,Srivastava2015prsa}
to determine the restrictions on Willis materials. In this section,
we extend the analysis to piezoelectric materials that exhibit electro-momentum
coupling, where by assuming passivity we derive restrictions on the
constitutive tensors given in Eqs.~\eqref{eq:EffConstRel} and$\ $\eqref{eq:EffConstRel-1}.

We consider a  piezoelectric solid of volume $\vol$ that is surrounded
by  air. Across its boundary $\bnd$ a surface charge density $\surfacech$
and traction $\traction$ are present, in addition to the volume densities
$\charge$ and $\force$. For simplicity, eigenstrains are not considered
here, bearing in mind that the effective properties to be used in
the sequel are those identified using such eigenstrains. Assuming
time-harmonic fields, we can express the complex rate of work done
on the piezoelectric body by the mechanical and electrical sources,
namely,
\begin{eqnarray}
\ComplexPower & = & \ointop_{\bnd}\left(\frac{\traction\cdot\dt{\disp}^{*}}{2}+\frac{\potential\dt{\surfacech}^{*}}{2}\right)\mathrm{d}a+\intvol{\left(\frac{\force\cdot\dt{\disp}^{*}}{2}+\frac{\potential\dt{\charge}^{*}}{2}\right)},\label{eq:ComplexP0}
\end{eqnarray}
such that the real part of $\ComplexPower$ is the time-average power
done by the sources \cite{auld1973acoustic}.  Using the connections
$\traction=\Stress\cdot\nh$ and $\Edisplacement\cdot\nh=-\surfacech,$
where $\nh$ is a unit vector in the outward normal direction to $\bnd$,
we obtain a restatement of the complex Poynting's theorem for piezoelectric
media in the settings of the quasi-electrostatic approximation as
\cite{auld1973acoustic}
\begin{eqnarray}
\ComplexPower & = & \ointop_{\bnd}\left(\frac{\Stress\cdot\dt{\disp}^{*}}{2}-\frac{\potential\dt{\Edisplacement}^{*}}{2}\right)\cdot\nh\mathrm{d}a+\intvol{\left(\frac{\force\cdot\dt{\disp}^{*}}{2}+\frac{\potential\dt{\charge}^{*}}{2}\right)}\nonumber \\
 & = & \intvol{\left(\frac{\Stress:\grad\dt{\disp}^{*}}{2}+\frac{\dt{\momentum}\cdot\dt{\disp}^{*}}{2}-\frac{\grad\potential\cdot\dt{\Edisplacement}^{*}}{2}\right)}.\label{eq:ComplexPower1}
\end{eqnarray}
In the process, we have applied the divergence theorem and used the
field equations \eqref{NewtonLaw} and \eqref{GaussLaw} after the
expansion of the divergence operator. The imaginary part of this
volume integral relates to the total stored energy within $\vol$
(elastic, kinetic and electric energy) and its real part is the time-average
power loss of the system.  Since a passive material cannot generate
energy, the inflow of power is always non-negative, and hence $\AveragePowerLoss:=\mathrm{Re}\ComplexPower$
is non-negative too, where here and throughout the text we use $^{\prime}$ and $^{\prime\prime}$ to denote the real and imaginary parts of any variable, respectively.
 This requirement imposes restrictions
on the permitted values of the constitutive tensors, when $\AveragePowerLoss$
is expressed using the generalized Willis relations. Invoking statistical
homogeneity and considering plane wave solutions (assuming they are valid), we employ form \eqref{eq:simplification}
to write the condition on $\AveragePowerLoss$ as
\begin{align}
\AveragePowerLoss & =\frac{1}{2}\mathrm{Re}\left\{ \int_{\vol}\left(\dt{\dispIndex}_{i,j}^{*}\FTrans{\elascalar}_{ijkl}\dispIndex_{k,l}+\dt{\dispIndex}_{i,j}^{*}\adjoint{\FTrans{\Piezoelectricscalar}_{ijk}}\potential_{,k}+\dt{\dispIndex}_{i,j}^{*}\FTrans{\williscalar}_{ijk}\dt{\dispIndex}_{k}-\right.\right.\nonumber \\
 & \left.\qquad\qquad\qquad\potential_{,i}^{*}\FTrans{\Piezoelectricscalar}_{ikl}\dt{\dispIndex}_{k,l}+\potential_{,i}^{*}\FTrans{\Permittivityscalar}_{ik}\dt{\potential}_{,k}-\potential_{,i}^{*}\FTrans{\rgscalar}_{ik}\dtt{\dispIndex}_{k}+\right.\label{eq:DissPower3}\\
 & \left.\qquad\qquad\qquad\left.\dt{\dispIndex}_{i}^{*}\adjoint{\FTrans{\rgscalar}_{ik}}\dt{\potential}_{,k}+\dt{\dispIndex}_{i}^{*}\adjoint{\FTrans{\williscalar}_{ikl}}\dt{\dispIndex}_{k,l}+\dt{\dispIndex}_{i}^{*}\FTrans{\rho}_{ik}\dtt{\dispIndex}_{k}\right)\mathrm{d}\sys\right\} \geq0;\nonumber
\end{align}
here, we used the fact that $\mathrm{\mathrm{Re}\left\{ \grad\potential\cdot\dt{\Edisplacement}^{*}\right\} =Re}\left\{ \grad\potential^{*}\cdot\dt{\Edisplacement}\right\} $.
The components of $\FTrans{\lmat}$ appearing in Eq.$\ $\eqref{eq:DissPower3}
are the transforms at $\left(-\Wavevector,\omega\right)$, and we
note that by linearity there is no loss of generality when considering
a single $\Wavevector$ vector. Eq.$\ $\eqref{eq:DissPower3} is
simplified using the following relations. First, we introduce the
skew-Hermitian parts of $\FTrans{\boldsymbol{\rho}},\FTrans{\Permittivity},$
and $\FTrans{\elas}$, namely,
\begin{eqnarray}
\nhpart{\FTrans{\rho}_{ik}}=\frac{1}{2}\left(\FTrans{\rho}_{ik}-\FTrans{\rho}_{ki}^{*}\right), & \nhpart{\FTrans{\Permittivityscalar}_{ik}}=\frac{1}{2}\left(\FTrans{\Permittivityscalar}_{ik}-\FTrans{\Permittivityscalar}_{ki}^{*}\right),\nonumber \\
 & \nhpart{\FTrans{\elascalar}_{ijkl}}=\frac{1}{2}\left(\FTrans{\elascalar}_{ijkl}-\FTrans{\elascalar}_{klij}^{*}\right),\label{eq:skew}
\end{eqnarray}
 to rewrite the terms $\mathrm{Re}\left\{ \dt{\dispIndex}_{i}^{*}\FTrans{\rho}_{ik}\dtt{\dispIndex}_{k}\right\} ,\mathrm{Re}\left\{ \potential_{,i}^{*}\FTrans{\Permittivityscalar}_{ik}\dt{\potential}_{,k}\right\} $,
and $\mathrm{Re}\left\{ \dt{\dispIndex}_{i,j}^{*}\FTrans{\elascalar}_{ijkl}\dispIndex_{k,l}\right\} $
as \begin{subequations}\label{eqs:Realterms}
\begin{eqnarray}\label{eq:ReDensity}
\mathrm{Re}\left\{ \dt{\dispIndex}_{i}^{*}\FTrans{\rho}_{ik}\dtt{\dispIndex}_{k}\right\} &=&\frac{1}{2}\left(\dt{\dispIndex}_{i}^{*}\FTrans{\rho}_{ik}\dtt{\dispIndex}_{k}+\dt{\dispIndex}_{i}\FTrans{\rho}_{ik}^{*}\dtt{\dispIndex}_{k}^{*}\right)\nonumber\\
&=&-i\omega\nhpart{\FTrans{\rho}_{ik}}\dt{\dispIndex}_{i}^{*}\dt{\dispIndex}_{k},
    \end{eqnarray}
\begin{eqnarray}\label{eq:RePerm}
\mathrm{Re}\left\{ \potential_{,i}^{*}\FTrans{\Permittivityscalar}_{ik}\dt{\potential}_{,k}\right\} &=&\frac{1}{2}\left(\potential_{,i}^{*}\FTrans{\Permittivityscalar}_{ik}\dt{\potential}_{,k}+\potential_{,i}\FTrans{\Permittivityscalar}_{ik}^{*}\dt{\potential}_{,k}^{*}\right)\nonumber\\
&=&-i\omega\nhpart{\FTrans{\Permittivityscalar}_{ik}}\potential_{,i}^{*}\potential_{,k},
    \end{eqnarray}
\begin{eqnarray}\label{eq:ReStiff}
\mathrm{Re}\left\{ \dt{\dispIndex}_{i,j}^{*}\FTrans{\elascalar}_{ijkl}\dispIndex_{k,l}\right\} &=&\frac{1}{2}\left(\dt{\dispIndex}_{i,j}^{*}\FTrans{\elascalar}_{ijkl}\dispIndex_{k,l}+\dt{\dispIndex}_{i,j}\FTrans{\elascalar}_{ijkl}^{*}\dispIndex_{k,l}^{*}\right)\nonumber\\
&=&i\omega\nhpart{\FTrans{\elascalar}_{ijkl}}\dispIndex_{i,j}^{*}\dispIndex_{k,l}.
    \end{eqnarray}
\end{subequations} We also note that \emph{if}
\begin{equation}
\FTrans{\rho}_{ik}=\FTrans{\rho}_{ki},\quad,\FTrans{\Permittivityscalar}_{ik}=\FTrans{\Permittivityscalar}_{ki},\quad\FTrans{\elascalar}_{ijkl}=\FTrans{\elascalar}_{klij},\label{eq:willis local reciprocity-1}
\end{equation}
\emph{for all} $\Wavevector$, then their skew-Hermitian part is equal
to their imaginary part (and the Hermitian part is equal to the real
part). The remaining terms can be written as

\begin{widetext}\begin{subequations}\label{eqs:Realterms2}
\begin{eqnarray}\label{eq:RePiezo}
\mathrm{Re}\left\{ \dt{\dispIndex}_{i,j}^{*}\adjoint{\FTrans{\Piezoelectricscalar}_{ijk}}\potential_{,k}-\potential_{,i}^{*}\FTrans{\Piezoelectricscalar}_{ikl}\dt{\dispIndex}_{k,l}\right\}
&=&\frac{i\omega}{2}\left(\adjoint{\FTrans{\Piezoelectricscalar}_{ijk}}-\FTrans{\Piezoelectricscalar}_{kij}^{*}\right)\potential_{,k}\dispIndex_{i,j}^{*}-\frac{i\omega}{2}\left(\FTrans{\Piezoelectricscalar}_{ijk}^{\dagger*}-\FTrans{\Piezoelectricscalar}_{kij}\right)\potential_{,k}^{*}\dispIndex_{i,j} ,\nonumber\\
&=&\omega\mathrm{Re}\left\{ i\left(\adjoint{\FTrans{\Piezoelectricscalar}_{ijk}}-\FTrans{\Piezoelectricscalar}_{kij}^{*}\right)\potential_{,k}\dispIndex_{i,j}^{*}\right\}\eqqcolon 2\omega\mathrm{Re}\left\{ i\potential_{,k}\qnhpart{\FTrans{\Piezoelectricscalar}}_{kij}\dispIndex_{i,j}^{*}\right\},
\end{eqnarray}
\begin{eqnarray}\label{eq:ReWillis}
\mathrm{Re}\left\{ \dt{\dispIndex}_{i,j}^{*}\FTrans{\williscalar}_{ijk}\dt{\dispIndex}_{k}+\dt{\dispIndex}_{i}^{*}\adjoint{\FTrans{\williscalar}_{ikl}}\dt{\dispIndex}_{k,l}\right\}
&=&\frac{i\omega}{2}\left(\FTrans{\williscalar}_{ijk}+\FTrans{\williscalar}_{kij}^{\dagger*}\right)\dt{\dispIndex}_{k}\dispIndex_{i,j}^{*}-\frac{i\omega}{2}\left(\FTrans{\williscalar}_{ijk}^{*}+\FTrans{\williscalar}_{kij}^{\dagger}\right)\dt{\dispIndex}_{k}^{*}\dispIndex_{i,j},\nonumber\\
&=& \omega\mathrm{Re}\left\{ -i\left(\FTrans{\williscalar}_{ijk}^{*}+\FTrans{\williscalar}_{kij}^{\dagger}\right)\dt{\dispIndex}_{k}^{*}\dispIndex_{i,j}\right\}\eqqcolon 2\omega\mathrm{Re}\left\{ -i\dispIndex_{i,j}\qhpart{\FTrans{\williscalar}_{ijk}}\dt{\dispIndex}_{k}^{*}\right\},\nonumber\\
&=& \omega\mathrm{Re}\left\{ -i\left(\modsc_{ijk}^{*}-\adjoint{\modsc_{kij}}\right)\ddot{\dispIndex}_{k}^{*}\dispIndex_{i,j}\right\}\eqqcolon 2\omega\mathrm{Re}\left\{i\dispIndex_{i,j}\qnhpart{\modsc_{ijk}}\ddot{\dispIndex}_{k}^{*}\right\},
\end{eqnarray}
\begin{eqnarray}\label{eq:ReEMC}
\mathrm{Re}\left\{ \dt{\dispIndex}_{i}^{*}\adjoint{\FTrans{\rgscalar}_{ik}}\dt{\potential}_{,k}-\potential_{,i}^{*}\FTrans{\rgscalar}_{ik}\dt{\dispIndex}_{k}\right\} &=&\frac{i\omega}{2}\left(\FTrans{\rgscalar}_{ik}^{\dagger*}+\FTrans{\rgscalar}_{ki}\right)\dt{\dispIndex}_{i}\potential_{,k}^{*}-\frac{i\omega}{2}\left(\adjoint{\FTrans{\rgscalar}_{ik}}+\FTrans{\rgscalar}_{ki}^{*}\right)\dt{\dispIndex}_{i}^{*}\potential_{,k},\nonumber\\
&=&\omega\mathrm{Re}\left\{ -i\left(\FTrans{\rgscalar}_{ki}^{*}+\adjoint{\FTrans{\rgscalar}_{ik}}\right)\dt{\dispIndex}_{i}^{*}\potential_{,k}\right\}\eqqcolon 2\omega\mathrm{Re}\left\{ -i\potential_{,k}\qhpart{\FTrans{\rgscalar}_{ki}}\dt{\dispIndex}_{i}^{*}\right\}  ,\nonumber\\
&=&\omega\mathrm{Re}\left\{-i \left(\modwc_{ki}^{*}-\adjoint{\modwc_{ik}}\right)\ddot{\dispIndex}_{i}^{*}\potential_{,k}\right\}\eqqcolon 2\omega\mathrm{Re}\left\{i\potential_{,k}\qnhpart{\modwc_{ki}}\ddot{\dispIndex}_{i}^{*}\right\} .
\end{eqnarray}
\end{subequations}

\end{widetext}  \emph{If} the following symmetries hold
\begin{equation}
\adjoint{\FTrans{\Piezoelectricscalar}_{ijk}}=\FTrans{\Piezoelectricscalar}_{kij},\adjoint{\modsc_{kij}}=\modsc_{ijk},\adjoint{\modwc_{ik}}=\modwc_{ki}\label{eq:assumed symmetries}
\end{equation}
\emph{for all} $\Wavevector$ then $\qnhpart{\left(\circ\right)}$
is equivalent to the imaginary part of $\left(\circ\right)$, while
$\qhpart{\left(\circ\right)}$ is equivalent to the real part. As we show in Sec.\ \ref{sec:Reciprocity}, symmetries \eqref{eq:willis local reciprocity-1} and \eqref{eq:assumed symmetries} are compatible with reciprocity only in the-long wavelength limit.
Using relations \eqref{eqs:Realterms2}, Eq.~\eqref{eq:DissPower3} reads
\begin{align}
\AveragePowerLoss & =\frac{\omega}{2}\int_{\vol}\left(\dispIndex_{i,j}^{*}i\nhpart{\FTrans{\elascalar}_{ijkl}}\dispIndex_{k,l}+2\mathrm{Re}\left\{ \potential_{,k}i\qnhpart{\FTrans{\Piezoelectricscalar}_{kij}}\dispIndex_{i,j}^{*}\right\} \right.\nonumber \\
 & \qquad\qquad\left.-2\mathrm{Re}\left\{ \dispIndex_{i,j}i\qhpart{\FTrans{\williscalar}_{ijk}}\dt{\dispIndex}_{k}^{*}\right\} -\dt{\dispIndex}_{i}^{*}i\nhpart{\FTrans{\rho}_{ik}}\dt{\dispIndex}_{k}\right.\nonumber \\
 & \qquad\qquad\left.-\potential_{,i}^{*}i\nhpart{\FTrans{\Permittivityscalar}_{ik}}\potential_{,k}-2\mathrm{Re}\left\{ \potential_{,k}i\qhpart{\FTrans{\rgscalar}_{ki}}\dt{\dispIndex}_{i}^{*}\right\} \right)\mathrm{d}\sys\geq0.\label{eq:DissPower4}
\end{align}
To proceed, we follow the argument of \citet{Muhlestein20160Prsa2}, which requires restricting subsequent analysis to the long-wavelength limit ($\Wavevector=\mathbf{0}$). In this limiting case, the strain, velocity
and electric fields in  Eq.\ \eqref{eq:DissPower4} can be prescribed arbitrarily and independently of each other through suitable sets of (boundary and volume) sources. Accordingly, we can recover
first the conclusions of \citet{Srivastava2015prsa} and \citet{Muhlestein20160Prsa2}
in the limiting elastic case, by considering a configuration where
the electric field vanishes, for which
\begin{align}
\AveragePowerLoss & =\frac{\omega}{2}\int_{\vol}\left(\dispIndex_{i,j}^{*}i\nhpart{\FTrans{\elascalar}_{ijkl}}\dispIndex_{k,l}-2\mathrm{Re}\left\{ \dispIndex_{i,j}i\qhpart{\FTrans{\williscalar}_{ijk}}\dt{\dispIndex}_{k}^{*}\right\} \right.\nonumber \\
 & \qquad\qquad\left.-\dt{\dispIndex}_{i}^{*}i\nhpart{\FTrans{\rho}_{ik}}\dt{\dispIndex}_{k}\right)\mathrm{d}\sys\geq0.\label{eq:DissPower5}
\end{align}
and in terms of $\mods$
\begin{align}
\AveragePowerLoss & =\frac{\omega}{2}\int_{\vol}\left(\dispIndex_{i,j}^{*}i\nhpart{\FTrans{\elascalar}_{ijkl}}\dispIndex_{k,l}+2\mathrm{Re}\left\{ \dispIndex_{i,j}i\qnhpart{\modsc_{ijk}}\ddot{\dispIndex}_{k}^{*}\right\} \right.\nonumber \\
 & \qquad\qquad\left.-\dt{\dispIndex}_{i}^{*}i\nhpart{\FTrans{\rho}_{ik}}\dt{\dispIndex}_{k}\right)\mathrm{d}\sys\geq0.\label{eq:DissPowermods}
\end{align}
By setting the velocity to zero we obtain
\begin{equation}
\intvol{\dispIndex_{i,j}^{*}i\nhpart{\FTrans{\elascalar}_{ijkl}}\dispIndex_{k,l}}\geq0,\label{eq:ImC}
\end{equation}
where the case of a vanishing strain provides
\begin{equation}
\intvol{\dt{\dispIndex}_{i}^{*}i\nhpart{\FTrans{\rho}_{ik}}\dt{\dispIndex}_{k}}\leq0.\label{eq:ImRho}
\end{equation}
Eqs.~\eqref{eq:ImC}-\eqref{eq:ImRho} hold for arbitrary strain
and velocity fields if and only if the Hermitian \footnote{The product of $i$ and a skew-Hermitian form is Hermitian.}
forms $i\nhpart{\FTrans{\elas}}$ and $i\nhpart{\FTrans{\boldsymbol{\rho}}}$
are positive- and negative-definite, respectively. If the medium is
not only passive but also lossless, then the inequalities become equalities
which imply that $\FTrans{\elas}$ and $\FTrans{\boldsymbol{\rho}}$
are Hermitian; this agrees with the
notion that Hermiticity implies energy conservation \cite{Chew2008,moiseyev2011book,shmuel2020prapplied}. The equalities further imply that $\qhpart{\check{\willis}}$ and $\qnhpart{\mods}$ are null.

As mentioned, this analysis recovers the results of \citet{Srivastava2015prsa}
and \citet{Muhlestein20160Prsa2} for Milton-Briane-Willis materials (i.e., local Willis materials). To develop the
restrictions on the couplings that arise in the electroelastic setting,
we first assume a combination of sources for which the only non-vanishing
field is the electric field. In this setting, Eq.~\eqref{eq:DissPower4}
provides
\begin{equation}
\intvol{\potential_{,i}^{*}i\nhpart{\FTrans{\Permittivityscalar}_{ik}}\potential_{,k}}\leq0.\label{eq:ImA}
\end{equation}
Since $\grad\potential$ is arbitrary, this condition holds if and
only if the Hermitian form $i\nhpart{\FTrans{\Permittivity}}$ is
negative-definite, and in the lossless case implies that $\FTrans{\Permittivity}$
is Hermitian, again, in agreement with the association of Hermiticity
with energy conservation. If only the velocity vanishes, we have that
\begin{equation}
\intvol{\left(\dispIndex_{i,j}^{*}i\nhpart{\FTrans{\elascalar}_{ijkl}}\dispIndex_{k,l}+2\mathrm{Re}\left\{ \potential_{,k}i\qnhpart{\FTrans{\Piezoelectricscalar}_{kij}}\dispIndex_{i,j}^{*}\right\} -\potential_{,i}^{*}i\nhpart{\FTrans{\Permittivityscalar}_{ik}}\potential_{,k}\right)}\geq0,
\end{equation}
If only the strain is zero
\begin{equation}
\intvol{\left(-\dt{\dispIndex}_{i}^{*}i\nhpart{\FTrans{\rho}_{ik}}\dt{\dispIndex}_{k}-\potential_{,i}^{*}i\nhpart{\FTrans{\Permittivityscalar}_{ik}}\potential_{,k}-2\mathrm{Re}\left\{ \potential_{,k}i\qhpart{\FTrans{\rgscalar}_{ki}}\dt{\dispIndex}_{i}^{*}\right\} \right)}\geq0,
\end{equation}
from which we obtain
\begin{eqnarray}
-2\mathrm{Re}\left\{ \potential_{,k}i\qnhpart{\FTrans{\Piezoelectricscalar}}_{kij}\dispIndex_{i,j}^{*}\right\}  & \leq & \dispIndex_{i,j}^{*}i\nhpart{\FTrans{\elascalar}_{ijkl}}\dispIndex_{k,l}-\potential_{,i}^{*}i\nhpart{\FTrans{\Permittivityscalar}_{ik}}\potential_{,k},\label{eq:ImB}\\
2\mathrm{Re}\left\{ \potential_{,k}i\qhpart{\FTrans{\rgscalar}_{ki}}\dt{\dispIndex}_{i}^{*}\right\}  & \leq & -\dt{\dispIndex}_{i}^{*}i\nhpart{\FTrans{\rho}_{ik}}\dt{\dispIndex}_{k}-\potential_{,i}^{*}i\nhpart{\FTrans{\Permittivityscalar}_{ik}}\potential_{,k},\label{eq:ReW}
\end{eqnarray}
and the latter is replaced by
\begin{equation}
-2\mathrm{Re}\left\{ \potential_{,k}i\qnhpart{\modwc_{ki}}\dtt{\dispIndex}_{i}^{*}\right\} \leq-\dt{\dispIndex}_{i}^{*}i\nhpart{\FTrans{\rho}_{ik}}\dt{\dispIndex}_{k}-\potential_{,i}^{*}i\nhpart{\FTrans{\Permittivityscalar}_{ik}}\potential_{,k},\label{eq:imwmod}
\end{equation}
when expressed in terms of $\modw$. Eqs.$\ $\eqref{eq:ImB}-\eqref{eq:imwmod}
thus provide bounds for $\qnhpart{\FTrans{\Piezoelectricmodule}},\qhpart{\text{\ensuremath{\check{\rg}}}}$
and $\text{\ensuremath{\qnhpart{\modw}}}$, and in the lossless
case imply they are null.

\section{Reciprocity\label{sec:Reciprocity}}

\begin{figure*}[t!]
\centering
\includegraphics[width=0.7\textwidth]{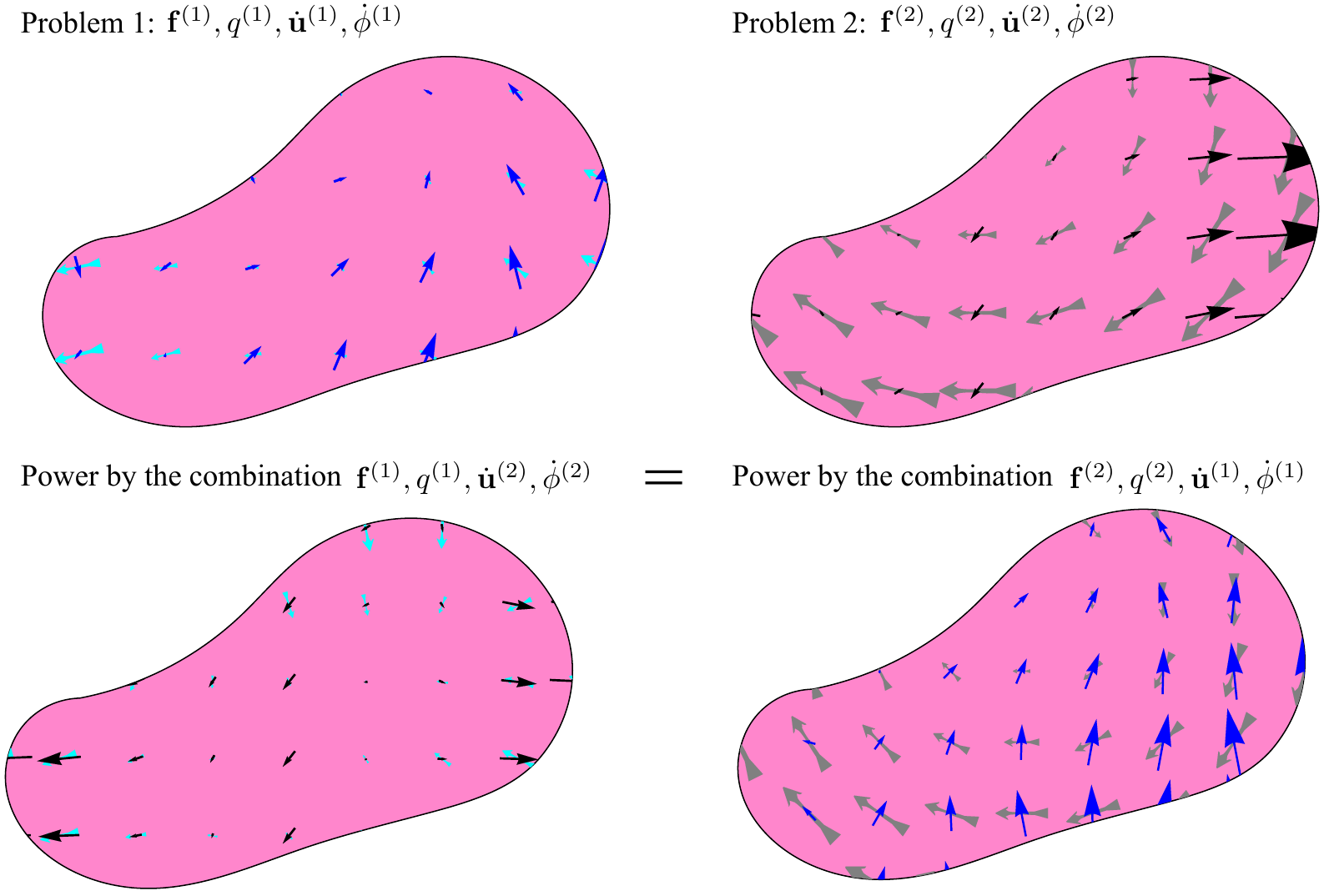}\caption{\label{fig:reciprocity} Schematics of the reciprocity principle. The source distributions of problems $1$ (cyan) and $2$ (gray) are illustrated by  darts. The resultant rates of the displacement and electric potential fields  of problems $1$ (blue) and $2$ (black) are illustrated by arrows. The body is
reciprocal if the power that distribution $1$ does along the fields
excited by distribution $2$ (bottom left sketch) is equal to the power that distribution
$2$ does along the fields excited by distribution $1$ (bottom right sketch).}
\end{figure*}

Consider a time-invariant piezoelectric body, two arbitrary  time-harmonic
source distributions, and denote these sources and the fields they
excite by superscripts $1$ and $2$, respectively.  The body is
reciprocal if the power that distribution $1$ does along the fields
excited by distribution $2$ is equal to the power that distribution
$2$ does along the fields excited by distribution $1$. A schematic illustration of this property is given in Fig.~\ref{fig:reciprocity}.

The principle
of reciprocity is independent of the level of isotropy and homogeneity
of the body \footnote{The real form employed next applies also if the material is active
or lossy, however such materials do not satisfy the complex reciprocity
relation \cite{auld1973acoustic}, unless gain and loss are balanced.}, however it requires that at each point the symmetry conditions
\begin{equation}
\Permittivityscalar_{ij}\left(\sys\right)=\Permittivityscalar_{ji}\left(\sys\right),\quad\elascalar_{ijkl}\left(\sys\right)=\elascalar_{klij}\left(\sys\right)\label{eq:major symmetries}
\end{equation}
are satisfied \cite{Achenbach2003book}. In the homogenization process
of a heterogeneous reciprocal body, it is thus required that the resultant
effective properties will also satisfy the reciprocity relation. \citet{Muhlestein20160Prsa2}
have shown that this requirement imposes the following conditions
on the effective properties of (spatially) \emph{local} Willis materials
\begin{equation}
\FTrans{\rho}_{ik}=\FTrans{\rho}_{ki},\quad\adjoint{\FTrans{\williscalar}_{ijk}}=\FTrans{\williscalar}_{jki},\quad\FTrans{\elascalar}_{ijkl}=\FTrans{\elascalar}_{klij},\label{eq:willis local reciprocity}
\end{equation}
where the symmetry between $\adjoint{\FTrans{\willis}}$ and $\FTrans{\willis}$
is transmitted to the modified couplings, namely,
\begin{equation}
\adjoint{\modsc_{ijk}}=\modsc_{jki}.\label{eq:sym modified S}
\end{equation}
 Here, we first derive the generalization of these conditions to
local materials exhibiting the piezo-momentum coupling, and then analyze
the general (non-local) case.  Our departure point towards this end
is the equations that govern the response of the body when subjected
to two independent and arbitrary distributions of force $\left\{ \forceindex_{i}^{(1)},\forceindex_{i}^{(2)}\right\} $
and charge densities $\left\{ \charge^{\left(1\right)},\charge^{\left(2\right)}\right\} $,
namely,
\begin{equation}
\left(\begin{array}{c}
\stressindex_{ij,j}^{(1)}-\dt{\scalarmomentum}_{i}^{(1)}\\
\Edisplaindex_{j,j}^{(1)}
\end{array}\right)+\left(\begin{array}{c}
\forceindex_{i}^{(1)}\\
-\charge^{(1)}
\end{array}\right)=\left(\begin{array}{c}
0_{i}\\
0
\end{array}\right),\label{eq:GovEq-Dist1}
\end{equation}
and
\begin{equation}
\left(\begin{array}{c}
\stressindex_{ij,j}^{(2)}-\dt{\scalarmomentum}_{i}^{(2)}\\
\Edisplaindex_{j,j}^{(2)}
\end{array}\right)+\left(\begin{array}{c}
\forceindex_{i}^{(2)}\\
-\charge^{(2)}
\end{array}\right)=\left(\begin{array}{c}
0_{i}\\
0
\end{array}\right).\label{eq:GovEq-Dist2}
\end{equation}
We denote the resultant displacement and electric potential fields
by $\genpot^{\left(1\right)\mathsf{T}}\coloneqq\left\{ \dispIndex_{i}^{\left(1\right)},\potential^{\left(1\right)}\right\} $
and $\genpot^{\left(2\right)\mathsf{T}}\coloneqq\left\{ \dispIndex_{i}^{\left(2\right)},\potential^{\left(2\right)}\right\} $,
respectively. Next, we left-multiply Eqs.~\eqref{eq:GovEq-Dist1}
and \eqref{eq:GovEq-Dist2} by $\dot{\genpot}^{\left(2\right)\mathsf{T}}$
and $\dot{\genpot}^{\left(1\right)\mathsf{T}}$, respectively. The
difference between the two products is
\begin{equation}
\begin{aligned}\stressindex_{ij,j}^{(1)}\dt{\dispIndex}_{i}^{(2)}+\dt{\scalarmomentum}_{i}^{(2)}\dt{\dispIndex}_{i}^{(1)}+\dt{\dispIndex}_{i}^{(2)}\forceindex_{i}^{(1)}+\dt{\potential}^{(2)}\Edisplaindex_{j,j}^{(1)}+\dt{\potential}^{(1)}\charge^{(2)}\ \\
-\left(\stressindex_{ij,j}^{(2)}\dt{\dispIndex}_{i}^{(1)}+\dt{\scalarmomentum}_{i}^{(1)}\dt{\dispIndex}_{i}^{(2)}+\dt{\dispIndex}_{i}^{(1)}\forceindex_{i}^{(2)}+\dt{\potential}^{(1)}\Edisplaindex_{j,j}^{(2)}+\dt{\potential}^{(2)}\charge^{(1)}\right) & =0,
\end{aligned}
\label{eq:ReactionExp1}
\end{equation}
which can rearranged as
\begin{align}
\dt{\scalarmomentum}_{i}^{(1)}\dt{\dispIndex}_{i}^{(2)}-\dt{\scalarmomentum}_{i}^{(2)}\dt{\dispIndex}_{i}^{(1)}+\stressindex_{ij}^{(1)}\dt{\dispIndex}_{i,j}^{(2)}-\stressindex_{ij}^{(2)}\dt{\dispIndex}_{i,j}^{(1)}\nonumber \\
+\dt{\potential}_{,j}^{(2)}\Edisplaindex_{j}^{(1)}-\dt{\potential}_{,j}^{(1)}\Edisplaindex_{j}^{(2)} & =\recip,\label{eq:ReactExp2}
\end{align}
 using the identities \begin{subequations}\label{eqs:minorSym1}
\begin{eqnarray}\label{eq:div1}
\stressindex_{ij,j}^{(1)}\dt{\dispIndex}_{i}^{(2)}-\stressindex_{ij,j}^{(2)}\dt{\dispIndex}_{i}^{(1)}&=&\left\{ \stressindex_{ij}^{(1)}\dt{\dispIndex}_{i}^{(2)}-\stressindex_{ij}^{(2)}\dt{\dispIndex}_{i}^{(1)}\right\} {}_{,j}\nonumber\\&&
-\left(\stressindex_{ij}^{(1)}\dt{\dispIndex}_{i,j}^{(2)}-\stressindex_{ij}^{(2)}\dt{\dispIndex}_{i,j}^{(1)}\right),
\end{eqnarray}
\begin{eqnarray}\label{eq:div2}
\dt{\potential}^{(2)}\Edisplaindex_{j,j}^{(1)}-\dt{\potential}^{(1)}\Edisplaindex_{j,j}^{(2)}&=&\left\{ \dt{\potential}^{(2)}\Edisplaindex_{j}^{(1)}-\dt{\potential}^{(1)}\Edisplaindex_{j}^{(2)}\right\} {}_{,j}\nonumber\\
&&-\left(\dt{\potential}_{,j}^{(2)}\Edisplaindex_{j}^{(1)}-\dt{\potential}_{,j}^{(1)}\Edisplaindex_{j}^{(2)}\right),
\end{eqnarray}
\end{subequations}where
\begin{equation}
\begin{aligned}\recip= & \left\{ \stressindex_{ij}^{(1)}\dt{\dispIndex}_{i}^{(2)}+\Edisplaindex_{j}^{(1)}\dt{\potential}^{(2)}\right\} {}_{,j}+\forceindex_{i}^{(1)}\dt{\dispIndex}_{i}^{(2)}-\charge^{(1)}\dt{\potential}^{(2)}-\\
 & \left\{ \stressindex_{ij}^{(2)}\dt{\dispIndex}_{i}^{(1)}+\Edisplaindex_{j}^{(2)}\dt{\potential}^{(1)}\right\} {}_{,j}-\left(\forceindex_{i}^{(2)}\dt{\dispIndex}_{i}^{(1)}-\charge^{(2)}\dt{\potential}^{(1)}\right).
\end{aligned}
\label{eq:Qexp}
\end{equation}
The term $\recip$ is the differential form of the difference between
the power that distribution $1$ does along the fields excited by
distribution $2$ and the power that distribution $2$ does along
the fields excited by distribution $1$, hence vanishes if the body
is reciprocal. The global form is obtained by volume integration,
conversion of the first and third terms in the integral into surface
integrals using the divergence theorem, and identification of the
boundary sources $\load t_{i}=\load{\stressindex_{ij}}n_{j}$ and
$-\load{\surfacech}=\load{\Edisplaindex_{j}}n_{j}$ of distribution
$\dist$.

We now expand the terms on the left side of Eq.$\ $\eqref{eq:ReactExp2}
using the effective constitutive equations \eqref{eq:simplification}
in their spatially local form $\left(\Wavevector=\mathbf{0}\right)$
to obtain
\begin{equation}
\begin{aligned} & \ \left(\FTrans{\williscalar}_{ijk}-\adjoint{\FTrans{\williscalar}_{kij}}\right)\left(\dt{\dispIndex}_{k}^{(1)}\dispIndex_{i,j}^{(2)}-\dt{\dispIndex}_{k}^{(2)}\dispIndex_{i,j}^{(1)}\right)+\left(\FTrans{\rho}_{ik}-\FTrans{\rho}_{ki}\right)\dt{\dispIndex}_{i}^{(2)}\dt{\dispIndex}_{k}^{(1)}\\
 & +\left(\FTrans{\rgscalar}_{ik}-\adjoint{\FTrans{\rgscalar}_{ki}}\right)\left(\dt{\dispIndex}_{k}^{(1)}\potential_{,i}^{(2)}-\dt{\dispIndex}_{k}^{(2)}\potential_{,i}^{1)}\right)+\left(\FTrans{\Permittivityscalar}_{ik}-\FTrans{\Permittivityscalar}_{ki}\right)\potential_{,k}^{(2)}\potential_{,i}^{(1)}\\
 & +\left(\adjoint{\FTrans{\Piezoelectricscalar}_{ijk}}-\FTrans{\Piezoelectricscalar}_{kij}\right)\left(\potential_{,k}^{(1)}\dispIndex_{i,j}^{(2)}-\potential_{,k}^{(2)}\dispIndex_{i,j}^{(1)}\right)\\
 & +\left(\FTrans{\elascalar}_{ijkl}-\FTrans{\elascalar}_{klij}\right)\dispIndex_{k,l}^{(1)}\dispIndex_{i,j}^{(2)}=0.
\end{aligned}
\label{eq:ReacExp-3}
\end{equation}
The arbitrariness of the sources implies that the strain, electric
and velocity fields are arbitrary too. Accordingly, for Eq.$\ $\eqref{eq:ReacExp-3}
to hold for any $\grad\disp,\dot{\disp},$ and $\grad\potential$,
the effective constitutive tensors must satisfy
\begin{equation}
\FTrans{\Permittivityscalar}_{ik}=\FTrans{\Permittivityscalar}_{ki},\quad\adjoint{\FTrans{\Piezoelectricscalar}_{ijk}}=\FTrans{\Piezoelectricscalar}_{kij},\quad\adjoint{\FTrans{\rgscalar}_{ki}}=\FTrans{\rgscalar}_{ik},\label{eq:rg local reciprocity}
\end{equation}
in addition to restrictions \eqref{eq:willis local reciprocity}.
It is clear that the modified couplings $\adjoint{\modw}$ and $\modw$
exhibit the same symmetry between $\adjoint{\FTrans{\rg}}$ and $\FTrans{\rg}$,
such that
\begin{equation}
\adjoint{\modwc_{ki}}=\modwc_{ik}.\label{eq:sym modified S-1}
\end{equation}
In view of Eqs.$\ $\eqref{eq:willis local reciprocity} and \eqref{eq:rg local reciprocity},
we can now revisit the conclusions in Sec.$\ $\ref{sec:Reciprocity-Passivity}
and replace the conditions on the Hermitian and skew-Hermitian parts
of the tensors \emph{in the long-wavelength limit} by the conditions
on their real and imaginary parts, respectively.

The foregoing analysis was obtained in the long-wavelength limit.
We derive next the general result for arbitrary wavelengths, and show
 that restrictions \eqref{eq:willis local reciprocity} and \eqref{eq:rg local reciprocity}
are its specialization. This will be carried out by showing that if
the body is reciprocal, then the governing equations are self-adjoint,
and in turn so is the Green function, which renders the constitutive
operator $\lop$ self-adjoint too. The latter property was remarked
only in passing by \citet{Willis2011PRSA,WILLIS2012MRC} and \citet{PernasSalomon2019JMPS}
in their respective problems, perhaps because the notion that reciprocity
and self-adjointness are closely related is somewhat known \cite{Knopoff1959,Gangi2000}.
However, since the self-adjoint structure of the constitutive operator
clearly depends on the definition of the effective description, it
is discussed in more detail here.

To proceed, it is useful to employ the formulation of \citet{Barnett1975},
who formulated the piezoelectric problem in a generalized space using
the following definitions
\begin{equation}
\begin{array}{ccc}
\genkcomp_{\alpha i\beta j} & = & \begin{cases}
\elascalar_{\alpha i\beta j} & \alpha,\beta\in\left\{ 1,2,3\right\} ,\\
\tp{\Piezoelectricscalar_{\alpha ij}} & \beta=4,\ \alpha\in\left\{ 1,2,3\right\} ,\\
\Piezoelectricscalar_{i\beta j} & \alpha=4,\ \beta\in\left\{ 1,2,3\right\} ,\\
-\Permittivityscalar_{ij} & \alpha=\beta=4,
\end{cases}\\
\gend_{\alpha\beta} & = & \begin{cases}
\delta_{\alpha\beta} & \alpha,\beta\in\left\{ 1,2,3\right\} ,\\
0 & \alpha\ \mathrm{or}\ \beta=4,
\end{cases}\\
\genb_{\alpha} & = & \begin{cases}
\scalarf_{\alpha} & \alpha\in\left\{ 1,2,3\right\} ,\\
-\charge & \alpha=4,
\end{cases}
\end{array}\label{eq:genfields}
\end{equation}
where the range of latin subscripts is limited to $\left\{ 1,2,3\right\} $.
The unified governing equations in terms of $\mathbf{\genkcomp},\Lambda$
and $\mathbf{\genb}$ read in index notation
\begin{equation}
\left\{ \genkcomp_{\alpha i\beta j}\gendiscomp_{\beta,j}\right\} _{,i}+\rho\omega^{2}\gend_{\alpha\beta}\gendiscomp_{\beta}=-\genb_{\alpha},\label{eq:unified governing equations}
\end{equation}
which define the components $G_{\beta\gamma}\left(\sys,\dummyb\right)$
of the Green matrix via
\begin{equation}
\left\{ \genkcomp_{\alpha i\beta j}G{}_{\beta\gamma,j}\right\} _{,i}+\rho\omega^{2}\gend_{\alpha\beta}G_{\beta\gamma}=-\delta_{\alpha\gamma}\delta\left(\sys-\dummyb\right),\label{eq:green def}
\end{equation}
where $\delta\left(\sys-\dummyb\right)$ is the Dirac delta. Eq.$\ $\eqref{eq:green def}
spells out explicitly the components of the symbolic Eq.$\ $(9) by
\citet{PernasSalomon2019JMPS}. In the Appendix
we describe the standard procedure to obtain the adjoint equations
and corresponding adjoint Green tensor, and verify its components
satisfy
\begin{equation}
G_{\gamma\beta}^{\dagger}\left(\sys,\dummyb\right)=\conj G_{\beta\gamma}\left(\dummyb,\sys\right);\label{eq:adjoint green}
\end{equation}
we thus recover the known result that if the body satisfies $\conj{\rho}=\rho$
and $\adjoint{\genkcomp}_{\alpha i\beta j}=\genkcomp_{\alpha i\beta j}$,
where
\begin{equation}
\adjoint{\genkcomp}_{\alpha i\beta j}=\begin{cases}
\elasindex_{\alpha i\beta j}^{\mathsf{T}*} & \alpha,\beta\in\left\{ 1,2,3\right\} ,\\
\Piezoelectricindex_{\alpha ij}^{\mathsf{T}*} & \beta=4,\ \alpha\in\left\{ 1,2,3\right\} ,\\
\Piezoelectricscalar_{i\beta j}^{*} & \alpha=4,\ \beta\in\left\{ 1,2,3\right\} ,\\
-\Permittivityscalar_{ij}^{\mathsf{T}*} & \alpha=\beta=4,
\end{cases}\label{eq:adj k-1-1}
\end{equation}
which is the case by virtue of Eq.$\ $\eqref{eq:symmetries}, then
the piezoelectric problem is self-adjoint \cite{Yaghjian1973la,Yang2012vn}.
As explained in the Appendix, in this case $G_{\gamma\beta}^{\dagger}\left(\sys,\dummyb\right)=G_{\gamma\beta}\left(\sys,\dummyb\right)$,
which together with the previous result implies that
\begin{equation}
G_{\gamma\beta}\left(\sys,\dummyb\right)=\conj G_{\beta\gamma}\left(\dummyb,\sys\right).\label{eq:self adjoint G}
\end{equation}
We recall next the expression for the kernel $\effective{\lmat}$
obtained by \citet{PernasSalomon2019JMPS}, namely,
\begin{equation}
\effective{\lmat}=\ensemble{\lmat}-\ensemble{\lmat\gradt\tp{\left(\gradt\green\right)}\lmat}+\ensemble{\lmat\gradt\green^{\mathsf{T}}}\ensemble{\green}^{-\mathsf{T}}\ensemble{\tp{\left(\gradt\green\right)}\lmat};\label{eq:nonlocal L}
\end{equation}
the symbolic matrix formulation for $\effective{\lmat}$ translates
to the following components
\begin{equation}
\effective L_{\alpha i\beta j}=\begin{cases}
\effective{\elascalar}_{\alpha i\beta j} & \alpha,\beta\in\left\{ 1,2,3\right\} ,\\
\adjoint{\effective{\Piezoelectricscalar}_{\alpha ij}} & \alpha\in\left\{ 1,2,3\right\} ,\ \beta=4,\\
\effective{\Piezoelectricscalar}_{i\beta j} & \alpha=4,\ \beta\in\left\{ 1,2,3\right\} ,\\
-\effective{\Permittivityscalar}_{ij} & \alpha=\beta=4,\\
\effective{\williscalar}_{i\beta j} & \alpha\in\left\{ 1,2,3\right\} ,\ \beta=5,\\
\adjoint{\effective{\williscalar}}_{\alpha ij} & \alpha=5,\ \beta\in\left\{ 1,2,3\right\} ,\\
\effective{\rgscalar}_{ij} & \alpha=4\ \beta=5,\\
\adjoint{\effective{\rgscalar}_{ij}} & \alpha=5,\ \beta=4,\\
\effective{\rho}_{ij} & \alpha=\beta=5.
\end{cases}\label{eq:L effective components}
\end{equation}
Inspecting the components of Eq.$\ $\eqref{eq:nonlocal L},
and employing the symmetries of $\green,\mathbf{K}$ and the fact
that $\lmat=\conj{\lmat}$, verify the symmetry
\begin{equation}
\effective L_{\alpha i\beta j}\left(\sys,\dummyb\right)=\effective L_{\beta j\alpha i}\left(\dummyb,\sys\right),\label{eq:L self adjoint}
\end{equation}
where terms associated with the conventional couplings $\tilde{\Permittivity},\effective{\Piezoelectricmodule},\bte,\effective{\elas},$
and $\effectivemass$ \emph{and} the modified couplings $\modw,\adjoint{\modw},\mods$
and $\adjoint{\mods}$ also satisfy
\begin{equation}
\effective L_{\alpha i\beta j}\left(\sys,\dummyb\right)=\conj{\effective L_{\beta j\alpha i}}\left(\dummyb,\sys\right),\label{eq:standard coupling}
\end{equation}
while the couplings of Willis type in their original form satisfy
\begin{eqnarray}
 & \effective L_{\alpha i\beta j}\left(\sys,\dummyb\right)=-\conj{\effective L_{\beta j\alpha i}}\left(\dummyb,\sys\right),\nonumber \\
 & \alpha\in\left\{ 1,2,3,4\right\} ,\ \beta=5\ \&\ \beta\in\left\{ 1,2,3,4\right\} ,\ \alpha=5.\label{eq:Willis type coupling}
\end{eqnarray}
It is important to note that symmetry \eqref{eq:L self adjoint}---which
delivers the self-adjoint property of $\effective{\lmat}$ as explained
later---originates from the symmetry $\lmat=\text{\ensuremath{\tp{\lmat}}}$,
and does not require the composite to be lossless; symmetries \eqref{eq:standard coupling}
and \eqref{eq:Willis type coupling} originate from the assumption
that $\lmat$ is also real.

Interestingly, the modified cross-couplings $\modw,\adjoint{\modw},\mods$
and $\adjoint{\mods}$ are related via same symmetry as the conventional
couplings, \emph{i.e}., Eq.$\ $\eqref{eq:standard coupling}. Combining
Eqs.$\ $\eqref{eq:L self adjoint}-\eqref{eq:Willis type coupling}
implies that the conventional couplings and the modified cross-couplings
are real, while those of Willis type are pure imaginary.  Together
with fact that according to Eq.$\ $\eqref{eq:green def}  the Green
tensor is an even function of $\omega$ \footnote{Assuming the composite has time-independent properties.},
this result implies that in the space-time domain $\tilde{\Permittivity},\effective{\Piezoelectricmodule},\bte,\effective{\elas},\effectivemass$
and $\mods,\adjoint{\mods},\modw$ and $\adjoint{\modw}$ are real---as
they should since they relate real physical quantities. By contrast,
in the space-time domain the cross-couplings $\effective{\willis},\adjoint{\effective{\willis}},\effective{\rg}$
and $\adjoint{\rg}$ are pure imaginary---an unphysical result. This
observation agrees with the analysis of \citet{Norrisrspa2011PRSA}
in the purely elastic case, who showed that $\effective{\elas}$ and
$\effectivemass$ are real in the space-time domain, while $\effective{\willis}$
and $\adjoint{\effective{\willis}}$ are pure imaginary.

For statically homogeneous media, we can employ the Fourier transform
with respect to the translation $\sys-\dummyb$, and write these symmetries
using indices in the transformed domain as\begin{subequations}
\label{eqs:symmetriesabcr}
\begin{align}
\label{eq:c}
\conj{\FTrans C_{klij}}\left(\Wavevector,\omega\right)&=\FTrans C_{ijkl}\left(\Wavevector,\omega\right)=\FTrans C_{klij}\left(-\Wavevector,\omega\right),\\
\conj{\FTrans B_{kij}}\left(\Wavevector,\omega\right)&=\adjoint{\FTrans B_{ijk}}\left(\Wavevector,\omega\right)=\FTrans B_{kij}\left(-\Wavevector,\omega\right),\\
\conj{\FTrans A_{ji}}\left(\Wavevector,\omega\right)&=\FTrans A_{ij}\left(\Wavevector,\omega\right)=\FTrans A_{ji}\left(-\Wavevector,\omega\right),\\
\conj{\FTrans{\rho}_{ji}}\left(\Wavevector,\omega\right)&=\FTrans{\rho}_{ij}\left(\Wavevector,\omega\right)=\FTrans{\rho}_{ji}\left(-\Wavevector,\omega\right),
\end{align}
\end{subequations}for terms associated with conventional couplings; the modified cross-couplings
terms satisfy the same form of symmetries, such that\begin{subequations}
\label{eqs:symmetriesmode}
\begin{align}
\conj{\modsc_{jki}}\left(\Wavevector,\omega\right)=\adjoint{\modsc{}_{ijk}}\left(\Wavevector,\omega\right)=\modsc_{jki}\left(-\Wavevector,\omega\right),\\
\conj{\modwc_{ji}}\left(\Wavevector,\omega\right)=\adjoint{\modwc{}_{ij}}\left(\Wavevector,\omega\right)=\modwc_{ji}\left(-\Wavevector,\omega\right),
\end{align}
\end{subequations}while when they are in their original form they satisfy \begin{subequations}
\label{eqs:symmetrieskw}
\begin{align}
-\conj{\FTrans{\williscalar}_{jki}}\left(\Wavevector,\omega\right)&=\adjoint{\FTrans{\williscalar}_{ijk}}\left(\Wavevector,\omega\right)=\FTrans{\williscalar}_{jki}\left(-\Wavevector,\omega\right),\\
\label{eq:FTrW}
-\conj{\FTrans{\rgscalar}_{ji}}\left(\Wavevector,\omega\right)&=\adjoint{\FTrans{\rgscalar}_{ij}}\left(\Wavevector,\omega\right)=\FTrans{\rgscalar}_{ji}\left(-\Wavevector,\omega\right).
\end{align}
\end{subequations}The symmetry between $\effective{\rg}$ and $\adjoint{\effective{\rg}}$
is shown in detail in the Appendix. It is clear
that in the limit $\Wavevector=\mathbf{0}$, the symmetries \eqref{eqs:symmetriesabcr}
and \eqref{eqs:symmetrieskw} recover symmetries \eqref{eq:willis local reciprocity}
and \eqref{eq:rg local reciprocity}. Eqs.$\ $\eqref{eqs:symmetriesabcr}-\eqref{eqs:symmetrieskw}
also endow the adjoint notion to the symbol $\dagger$ for the non-local
operators $\adjoint{\sop}$ and $\adjoint{\wop}$, since these symmetries
imply that \begin{subequations}
\label{eqs:adjoint justify}
\begin{align}
\intvol{\sop\left(\dt{\disp}\left(\dummyb\right)\right):\grad\disp\left(\sys\right)}&=\intvol{\dt{\disp}\left(\sys\right)\cdot\adjoint{\sop}\left(\grad\disp\left(\dummyb\right)\right)},\\
\intvol{\wop\left(\dt{\disp}\left(\dummyb\right)\right)\cdot\grad\potential\left(\sys\right)}&=\intvol{\dt{\disp}\left(\sys\right)\cdot\adjoint{\wop}\left(\grad\potential\left(\dummyb\right)\right)}.
\end{align}
\end{subequations}as well as for $\adjoint{\bop}$, which satisfies
\begin{equation}
\intvol{\bop\left(\grad\disp\left(\dummyb\right)\right)\cdot\grad\potential\left(\sys\right)}=\intvol{\grad\disp\left(\sys\right)\cdot\adjoint{\bop}\left(\grad\potential\left(\dummyb\right)\right)};\label{eq:badjointproof}
\end{equation}
the non-local operators $\rop,\aop,$ and $\cop$ are self-adjoint
in the sense above. At the cost of repetition, we clarify that the
equality between the middle- and most-right terms in Eqs.$\ $\eqref{eqs:symmetriesabcr}-\eqref{eqs:symmetrieskw}
was obtained by relying only on the fact that $\lmat=\tp{\lmat}$.
This property leads to a self-adjoint effective operator $\lop$
in the sense above, and specifically render $\adjoint{\sop}$ and
$\adjoint{\wop}$ (resp.$\ $$\adjoint{\modsop}$ and $\adjoint{\modwop}$)
the adjoints of $\sop$ and $\wop$ (resp.$\ $$\modsop$ and $\modwop$);
the same goes for $\adjoint{\bop}$ and $\bop$ via Eq.$\ $\eqref{eq:badjointproof}.
Otherwise, it would by justified to replace the $\dagger$ notation
to distinguish them from the adjoint operators.

We further clarify that the equality between the left- and middle
terms in Eqs.$\ $\eqref{eqs:symmetriesabcr}-\eqref{eqs:symmetrieskw}
relies only on the fact that in the frequency domain the properties
of the composite satisfy $\FTrans{\lmat}=\FTrans{\lmat}^{\mathsf{T}*}$,
or in other words they are Hermitian. This\emph{ does not} necessarily
imply that $\lmat$ is symmetric (although it can be), and immediately
satisfies Eqs.$\ $\eqref{eq:ImC}-\eqref{eq:ImA} as equalities. A case where $\FTrans{\lmat}=\FTrans{\lmat}^{\mathsf{T}*}$ and $\FTrans{\lmat}\neq\FTrans{\lmat}^{\mathsf{T}}$ corresponds to a non-reciprocal medium whose losses are compensated by the energy it generates, such that on average the material is lossless and passive  (no energy loss or gain).

\section{Causality \label{sec:Causality}}

The principle of causality states that an effect must follow its
cause. This principle implies the analyticity of the response functions
of linear systems and vice versa, namely, analyticity implies causality
\cite{Toll1956PhysRev,GrootMazur}. With the interpretation of the
constitutive properties of linear materials as response functions,
causality through analyticity provides relations between their real
and imaginary parts of their (time) Fourier transforms. These relations
were first obtained in electromagnetics for the permeability and permittivity
tensors, where they are known as the Kramers-Krönig relations \cite{Kronig1926xy,land&lifs60book,Nistad2008PRB,Silveirinha2011PhysRevB}.
This concept was later on applied in other branches of physics---and
specifically in mechanics---to obtain conditions on the pertinent
constitutive properties \cite{Srivastava2015prsa,Carcione2019ys}.
\citet{alu2011prb,Alu2011-PhysRevB} has shown that in certain cases
the bianisotropic tensor is essential for respecting causality in passive
media. (We clarify that the model Al{\`{u}} considered is local.)
Indeed, some of the electromagnetic homogenization schemes from which
this cross-coupling tensor is absent violate causality in such media \cite{Depine2004,efros2004pre,Simovski2009bc}.
 Analogously,  \citet{Sieck2017prb} recognized the need in Willis
coupling to satisfy causality by the effective constitutive properties
in elastodynamics. We clarify that We emphasize that the information from the Kramers-Krönig relations is limited for active media, since it is possible to realize anomalous responses for real frequencies (such as antiresonance) using suitable causal polynomials, see for example Ref.~\cite{Vasquez2009prl}. To get useful results, it is thus necessary to couple causality with passivity \footnote{Graeme Milton, private communication. See also Ref.~\cite{milton1997}, where bounds on the dispersion relations are obtained using causality in passive media. Remarkably, these bounds are for finite ranges of frequencies}.

In this Sec., we develop the restrictions placed by causality on the
effective properties \eqref{eq:ConstiRel-1}, \emph{i.e.}, when microscopically
the medium exhibits the intrinsic piezoelectric effect, and macroscopically
exhibits also the effective electro-momentum coupling. The framework
developed in Ref.$\ $\cite{PernasSalomon2019JMPS} constitutes a
platform to carry out this task with respect to the effective operator
$\effective{\lmat}$, which we recall is non-local both in space and
time. To facilitate the analysis, we here focus on the long-wavelength
limit $\Wavevector=\mathbf{0}$, and neglect spatially non-local effects
on causality \cite{Ringbauer2016kx}. Accordingly,  we omit the
spatial dependency of the fields in Eq.~\eqref{eq:Time-convolution},
and rewrite it as
\begin{equation}
\hmat\left(t\right)=\int_{-\infty}^{t}\effective{\lmat}\left(t-T\right)\left(T\right)\gmat\left(T\right)\mathrm{d}T,\label{eq:Linear}
\end{equation}
bearing in mind it holds at each material point. Let $\tau=t-T$;
 causality implies that $\effective{\lmat}$ must satisfy
\begin{equation}
\effective{\lmat}\left(\tau\right)=0\;\mathrm{for}\;\tau<0.\label{eq:Causalitycond}
\end{equation}
The approach taken here to relate the real and imaginary parts of
its Fourier transform is standard, and employs   the Plemelj formulas
see., \emph{e.g.}, Ref.$\ $\cite{Nussenzveig}. This approach is
summarized next to provide a self-contained analysis. As discussed
by \citet{Nussenzveig}, certain assumptions regarding $\KernelL\left(t\right)$
are required in order for $\FTrans{\lmat}^{\prime}\left(\omega\right)$
and $\FTrans{\lmat}^{\prime\prime}\left(\omega\right)$ to be related, where we recall that $^{\prime}$ and $^{\prime\prime}$ denote the real and imaginary parts of a variable, respectively.
In view of Eq.$\ $\eqref{eq:Causalitycond}, Eq.$\ $\eqref{eq:space time Fourier transform definition}
obtains the from
\begin{equation}
\FTrans{\lmat}\left(\omega\right)=\int_{0}^{\infty}\KernelL\left(\tau\right)e^{i\omega\tau}\mathrm{d}\tau,\label{eq:fourier transform half plane}
\end{equation}
where the integral is only over $\mathbb{R}_{+}$, implying that $\FTrans{\lmat}(\omega)$
has an analytic continuation in the upper half of the complex plane.
If we further assume at first that $\KernelL(t)$ is square integrable,
then through the Parseval-Plancherel theorem, we have that $\FTrans{\lmat}(\omega)$
is square integrable along any line in the upper half of the complex
plane that is parallel to the real axis, such that \cite{Nussenzveig}
\begin{equation}
\lim_{\alpha\rightarrow\pm\infty}\FTrans{\lmat}\left(\omega^{\prime}+i\omega^{\prime\prime}\right)=0,\;\;\;\omega^{\prime\prime}\geq0.\label{eq:InfiniteBehavior}
\end{equation}
This property is employed in the application of Cauchy's integral
formula to a closed curve $\Gamma$ about an arbitrary point $\oz$
in the upper half of the complex plane $\left(\oz^{\prime\prime}>0\right)$
\begin{equation}
\FTrans{\lmat}\left(\oz\right)=\frac{1}{2\pi i}\oint_{\Gamma}\mathrm{\frac{\FTrans{\lmat}\left(\mathit{\CompVar}\right)}{\mathit{\CompVar}-\oz}d\mathit{\CompVar}},\label{eq:cauhcy formula}
\end{equation}
in order to show it reduces to an integration along the real axis
\begin{equation}
\FTrans{\lmat}\left(\oz\right)=\frac{1}{2\pi i}\int_{-\infty}^{+\infty}\frac{\FTrans{\lmat}\left(z\right)}{z-\oz}\mathrm{d}z,\ z^{\prime\prime}=0.\label{eq:Modfunction}
\end{equation}
The case of real $\oz$ is obtained using a closed contour that avoids
$\oz$ by a semi-circle of radius $\epsilon$, and taking the limit
$\epsilon\rightarrow0$ to show that
\begin{equation}
\FTrans{\lmat}\left(\oz\right)=\frac{1}{i\pi}\mathrm{\dashint_{-\infty}^{\infty}\frac{\FTrans{\lmat}\left(\mathit{\CompVar}\right)}{\mathit{\CompVar}-\oz}d\mathit{\CompVar}},\ z^{\prime\prime}=\oz^{\prime\prime}=0,\label{eq:CauchyIntegral}
\end{equation}
where

\begin{equation}
\mathrm{\dashint_{-\infty}^{\infty}}=\lim_{\epsilon\rightarrow0}\left(\int_{-\infty}^{\omega-\epsilon}+\int_{\omega+\epsilon}^{\infty}\right)
\end{equation}
denotes the Cauchy's principal value.  The real and imaginary parts
of Eq.$\ $\eqref{eq:CauchyIntegral} provide the following relations

\begin{subequations}\label{eqs:KK}
\begin{equation}\label{eq:KKreal}
\FTrans{\lmat}^{\prime}\left(\omega\right)=\frac{1}{\pi}\mathrm{\dashint_{-\infty}^{\infty}}\frac{\FTrans{\lmat}^{\prime\prime}\left(\CompVar\right)}{\mathit{\CompVar}-\omega}\mathrm{d}\mathit{\CompVar},
\end{equation}
\begin{equation}\label{eq:KKimaginary}
\FTrans{\lmat}^{\prime\prime}\left(\omega\right)=-\frac{1}{\pi}\mathrm{\dashint_{-\infty}^{\infty}}\mathrm{\frac{\FTrans{\lmat}^{\prime}\left(\CompVar\right)}{\mathit{\mathit{\CompVar}}-\omega}\mathrm{d}\mathit{\CompVar}}
\end{equation}
\end{subequations}between the $\FTrans{\lmat}^{\prime}\left(\omega\right)$ and $\FTrans{\lmat}^{\prime\prime}\left(\omega\right)$
at any real frequency $\omega$. An alternative form of relations
\eqref{eqs:KK} is obtained using the fact that $\KernelL(t)$ is
real, and hence
\begin{equation}
\conj{\FTrans{\lmat}}\left(\omega\right)=\left(\int_{0}^{\infty}\KernelL\left(t\right)e^{i\omega t}\mathrm{d}t\right)^{*}=\int_{0}^{\infty}\KernelL\left(t\right)e^{-i\omega t}\mathrm{d}t=\FTrans{\lmat}\left(-\omega\right),
\end{equation}
leading to \begin{subequations}\label{eqs:oddevenL}
\begin{equation}\label{eq:even}
\FTrans{\lmat}^{\prime}\left(\omega\right)=\FTrans{\lmat}^{\prime}\left(-\omega\right),
\end{equation}
\begin{equation}\label{eq:odd}
\FTrans{\lmat}^{\prime\prime}\left(\omega\right)=-\FTrans{\lmat}^{\prime\prime}\left(-\omega\right).
\end{equation}
\end{subequations}Employing these symmetries leads to \begin{subequations}\label{eqs:KK2}
\begin{equation}\label{eq:real2}
\FTrans{\lmat}^{\prime}(\omega)=\frac{2}{\pi}\dashint_{0}^{\infty}\frac{\CompVar\FTrans{\lmat}^{\prime\prime}\left(\CompVar\right)}{\mathit{\CompVar}^{2}-\omega^{2}}\mathrm{d}\mathit{\CompVar},
\end{equation}
\begin{equation}\label{eq:imaginary2}
\FTrans{\lmat}^{\prime\prime}(\omega)=-\frac{2}{\pi}\dashint_{0}^{\infty}\mathrm{\frac{\omega\FTrans{\lmat}^{\prime}\left(\CompVar\right)}{\mathit{\CompVar}^{2}-\omega^{2}}d\mathit{\CompVar}.}
\end{equation}
\end{subequations}

We examine next the case where $\FTrans{\lmat}(\omega)$ is not square
integrable but a bounded function. In this case, property \eqref{eq:InfiniteBehavior}
that is needed to obtain Eq.\eqref{eq:Modfunction} no longer holds,
and consequently the relation between $\FTrans{\lmat}^{\prime}\left(\omega\right)$
and $\FTrans{\lmat}^{\prime\prime}\left(\omega\right)$ can be determined
only up to an arbitrary real constant. To determine this constant,
a knowledge of the value of $\FTrans{\lmat}(\omega)$ at some real
frequency is needed. Say $\FTrans{\lmat}(\omega)$ is differentiable
and known at $\oz$; then, we can repeat the procedure that led to
Eqs.$\ $\eqref{eqs:KK} and \eqref{eqs:KK2}, only now we replace
$\FTrans{\lmat}\left(\omega\right)$ by the function
\begin{equation}
\Funtion\left(\omega\right)\coloneqq\frac{\FTrans{\lmat}\left(\omega\right)-\FTrans{\lmat}\left(\omega_{0}\right)}{\omega-\omega_{0}},
\end{equation}
since it is bounded for $\omega\rightarrow\omega_{0}$, analytic in
the upper half of the complex plane\emph{ and square integrable}.
 If we further assume that $\oz\rightarrow\infty$, the end result
can be put in the form \begin{subequations}\label{eqs:KK3}
\begin{align}
\label{eq:real3}
\FTrans{\lmat}^{\prime}\left(\omega\right)&=\frac{2}{\pi}\dashint_{0}^{\infty}\frac{\mathit{\CompVar}\FTrans{\lmat}^{\prime\prime}\left(\CompVar\right)}{\mathit{\CompVar}^{2}-\omega^{2}}\mathrm{d}\CompVar+\FTrans{\lmat}^{\prime}\left(\infty\right),\\
\label{eq:imaginary3}
\FTrans{\lmat}^{\prime\prime}\left(\omega\right)&=-\frac{2}{\pi}\dashint_{0}^{\infty}\mathrm{\frac{\omega\FTrans{\lmat}^{\prime}\left(\CompVar\right)}{\mathit{\CompVar}^{2}-\omega^{2}}d\mathit{\CompVar},}
\end{align}
\end{subequations}cf.$\ $Eq.$\ $(4.7) in Ref.$\ $\cite{Muhlestein20160Prsa2}. Eq.$\ $\eqref{eqs:KK3}
should hold for all the tensor-valued (history) functions that comprise
$\FTrans{\lmat}\left(\omega\right)$,\emph{ i.e.}, $\tilde{\Permittivity},\effective{\Piezoelectricmodule},\adjoint{\tilde{\Piezoelectricmodule}},\effective{\elas},$
as well as $\effective{\willis},\effective{\rg},$ their adjoints
and their modified versions $\mods$ and $\modw$. This conclusion
thus generalizes previous works by requiring that the electro-momentum
tensor is also subjected to relations of the Kramer-Krönig type, as
expected. We recall that Sec.$\ $\ref{sec:Reciprocity} showed that
the Fourier transforms of the constitutive tensors of reciprocal,
passive and lossless media are all real, including the modified cross-couplings
$\mods,\adjoint{\mods},\modw$ and $\adjoint{\modw}$, and excluding
$\check{\willis},\adjoint{\effective{\willis}},\FTrans{\rg}$ and
$\adjoint{\FTrans{\rg}}$ which are pure imaginary. When this conclusion
is combined with conditions \eqref{eqs:KK3}, we obtain in the long-wavelength
limit of reciprocal, passive and lossless media that
\begin{eqnarray}
 & \FTrans{\Permittivity}\left(\omega\right)=\FTrans{\Permittivity}^{\prime}\left(\infty\right),\FTrans{\Piezoelectricmodule}\left(\omega\right)=\FTrans{\Piezoelectricmodule}^{\prime}\left(\infty\right),\adjoint{\FTrans{\Piezoelectricmodule}}\left(\omega\right)=\FTrans{\Piezoelectricmodule}^{\mathsf{\dagger}\prime}\left(\infty\right),\nonumber \\
 & \FTrans{\elas}\left(\omega\right)=\FTrans{\elas}^{\prime}\left(\infty\right),\FTrans{\boldsymbol{\rho}}\left(\omega\right)=\FTrans{\boldsymbol{\rho}}^{\prime}\left(\infty\right),\label{eq:casualitylimitcases}
\end{eqnarray}
and similarly for the modified cross-couplings $\mods,\adjoint{\mods},\modw$
and $\adjoint{\modw}$
\begin{eqnarray}
 & \mods\left(\omega\right)=\mods^{\prime}\left(\infty\right),\adjoint{\mods}\left(\omega\right)=\mods^{\dagger\prime}\left(\infty\right),\nonumber \\
 & \modw\left(\omega\right)=\modw^{\prime}\left(\infty\right),\modw^{\dagger}\left(\omega\right)=\modw^{\dagger\prime}\left(\infty\right),\label{eq:modifiedcausality}
\end{eqnarray}
where notably, the couplings of Willis type in their original representation
must be null. This is clear from Eq.$\ $\eqref{eq:imaginary3}, as
the integrand in the right side is identically zero since these couplings
are pure imaginary, which then implies that the left side---which
is their imaginary part---also vanishes.

\section{\label{sec:Closure}Closure}

Piezoelectric- and piezomagnetic materials exhibit intrinsic coupling
with non-mechanical fields. Recently, it was shown that the effective
response of composites made of such constituents exhibit additional
cross-couplings that are absent from the response of the constituents,
and are of Willis type \cite{PernasSalomon2019JMPS}. The recent development of such generalized Willis materials
comes with the question: What are the mathematical restrictions that
their constitutive relations should satisfy in order to respect passivity,
reciprocity and causality? In this paper, we have addressed this question by adapting standard methodologies used in electromagnetics, elastodynamics
and mathematics \cite{Nussenzveig,Chew2008,Srivastava2015prsa,Muhlestein20160Prsa2}.

We arrived at the following findings. From passivity we obtained several
inequality conditions on the skew-Hermitian and Hermitian parts of
the Fourier transforms of the effective properties. From reciprocity
we found certain symmetry and adjoint relations that the effective
operator satisfies. These conditions generalize the conditions in Refs.\ \cite{Srivastava2015prsa,Muhlestein20160Prsa2}
for the Milton-Briane-Willis equations (i.e., the long-wavelength limit of the Willis equations), not only by accounting for the electro-momentum coupling, but also for non-local interactions, leading to  wavevector-dependent conditions. Finally, from causality we obtain relations of the Kramer-Krönig
type between the real and imaginary parts of the operator in the frequency
domain. A summary of the mathematical restrictions are given in Tab.~\ref{tab:Tab-restrictions}. One implication that follows these restrictions is that the additional cross-couplings in the time domain are not with the electric field and velocity, but with their time derivative. This insight is analogous to the insights in Refs.\ \cite{Milton2007njp,nassar2015willis,Muhlestein20160Prsa2,Sieck2017prb}, suggesting acceleration-dependent stress and strain rate-dependent momentum formulation in the elastic case.

We conclude this paper by highlighting the applicational impact of our results. Our conclusions assess how energy is converted in such metamaterials, and in turn what is the efficiency that devices based on these cross-couplings can achieve. Indeed, the counterpart of our conclusions in the acoustic setting \cite{Alu2011-PhysRevB,Srivastava2015prsa,Muhlestein20160Prsa2} guided \citet{quan2018prl} in the design of metasurfaces with maximum Willis coupling for sound steering. Similarly, we expect this work to promote the design of devices which exploit the electro-momentum coupling to efficiently manipulate mechanical waves.

\begin{table*}
\begin{tabular}{cccc}
\toprule
Property & Real-space reciprocity & Fourier-space reciprocity & Passivity when $\Wavevector=\mathbf{0}$\tabularnewline
\midrule
\addlinespace[0.1cm]
Elasticity & $\effective{\elas}\left(\position,\dummyb\right)=\tp{\effective{\elas}}\left(\dummyb,\position\right)$ & $\FTrans{\elas}\left(\Wavevector\right)=\tp{\FTrans{\elas}}\left(-\Wavevector\right)$ & $i\nhpart{\FTrans{\elas}}$ positive definite\tabularnewline\addlinespace[0.1cm]
\addlinespace[0.1cm]
\addlinespace[0.1cm]
Mass density & $\effective{\boldsymbol{\rho}}\left(\position,\dummyb\right)=\tp{\effective{\boldsymbol{\rho}}}\left(\dummyb,\position\right)$ & $\FTrans{\boldsymbol{\rho}}\left(\Wavevector\right)=\tp{\FTrans{\boldsymbol{\rho}}}\left(-\Wavevector\right)$ & $i\nhpart{\FTrans{\boldsymbol{\rho}}}$ negative definite\tabularnewline\addlinespace[0.1cm]
\addlinespace[0.1cm]
\addlinespace[0.1cm]
Willis Coupling & $\adjoint{\effective{\willis}}\left(\position,\dummyb\right)=\tp{\effective{\willis}}\left(\dummyb,\position\right)$ & $\adjoint{\check{\willis}}\left(\Wavevector\right)=\tp{\check{\willis}}\left(-\Wavevector\right)$ & bound for $\qhpart{\check{\willis}}$\tabularnewline\addlinespace[0.1cm]
\addlinespace[0.1cm]
\addlinespace[0.1cm]
Modified Willis coupling & $\adjoint{\mods}\left(\position,\dummyb\right)=\tp{\mods}\left(\dummyb,\position\right)$ & $\adjoint{\mods}\left(\Wavevector\right)=\tp{\mods}\left(-\Wavevector\right)$ & bound for $\qnhpart{\mods}$\tabularnewline\addlinespace[0.1cm]
\addlinespace[0.1cm]
\addlinespace[0.1cm]
Permittivity & $\effective{\Permittivity}\left(\position,\dummyb\right)=\tp{\effective{\Permittivity}}\left(\dummyb,\position\right)$ & $\FTrans{\Permittivity}\left(\Wavevector\right)=\tp{\FTrans{\Permittivity}}\left(-\Wavevector\right)$ & $i\nhpart{\FTrans{\Permittivity}}$ negative definite\tabularnewline\addlinespace[0.1cm]
\addlinespace[0.1cm]
\addlinespace[0.1cm]
Piezoelectric coupling & $\adjoint{\effective{\Piezoelectricmodule}}\left(\position,\dummyb\right)=\tp{\effective{\Piezoelectricmodule}}\left(\dummyb,\position\right)$ & $\adjoint{\FTrans{\Piezoelectricmodule}}\left(\Wavevector\right)=\tp{\FTrans{\Piezoelectricmodule}}\left(-\Wavevector\right)$ & bound for $\qnhpart{\FTrans{\Piezoelectricmodule}}$\tabularnewline\addlinespace[0.1cm]
\addlinespace[0.1cm]
\addlinespace[0.1cm]
Electro-momentum coupling & $\adjoint{\FTrans{\rg}}\left(\position,\dummyb\right)=\tp{\FTrans{\rg}}\left(\dummyb,\position\right)$ & $\adjoint{\FTrans{\rg}}\left(\Wavevector\right)=\tp{\FTrans{\rg}}\left(-\Wavevector\right)$ & bound for $\qhpart{\FTrans{\rg}}$\tabularnewline\addlinespace[0.1cm]
\addlinespace[0.1cm]
\addlinespace[0.1cm]
Modified EM coupling & $\adjoint{\modw}\left(\position,\dummyb\right)=\tp{\modw}\left(\dummyb,\position\right)$ & $\adjoint{\modw}\left(\Wavevector\right)=\tp{\modw}\left(-\Wavevector\right)$ &  bound for $\qnhpart{\modw}$\tabularnewline\addlinespace[0.1cm]
\bottomrule
\addlinespace[0.1cm]
\end{tabular}

\caption{\label{tab:Tab-restrictions}The mathematical restrictions that the
effective description of piezoelectric composites with generalized
Willis couplings satisfy, owing to reciprocity and passivity. The
restrictions that result from causality are of the Kramer-Krönig type,
namely, $\protect\FTrans{\protect\lmat}^{\prime\prime}\left(\omega\right)=-\frac{2}{\pi}\dashint_{0}^{\infty}\mathrm{\frac{\omega\protect\FTrans{\protect\lmat}^{\prime}\left(\protect\CompVar\right)}{\mathit{\protect\CompVar}^{2}-\omega^{2}}d\mathit{\protect\CompVar}}$.}
\end{table*}

\section*{Acknowledgments}

We thank Graeme Milton for enriching discussions on non-locality and uniqueness, his insights on causality, and for useful references.
We thank Doron Shilo for his observation regarding the similarity with
the broken inversion symmetry at the atomic scale in piezoelectric
materials. This research was supported by the Israel Science Foundation,
funded by the Israel Academy of Sciences and Humanities (Grant no.$\ $2061/20),
the United States-Israel Binational Science Foundation (Grant no.~2014358),
and Ministry of Science and Technology (grant no.~880011).

\appendix

\global\long\def\theequation{A.\arabic{equation}}%
 \setcounter{equation}{0}

\section*{\label{Appendix green function}Appendix A. Derivations related to the adjoint operator and Green tensor}

The left side of Eq.$\ $\eqref{eq:unified governing equations} defines
the action of an operator $\constop$ on the vector field $\genpot$.
Accordingly, the left side of Eq.$\ $\eqref{eq:unified governing equations}
is the $\alpha$-component of the action of $\constop$ on the vector
field $G_{.\left(\gamma\right)}$, whose components are $G_{\beta\left(\gamma\right)},\beta=1,2,3,4.$
The adjoint operator $\adjoint{\constop}$ is defined via the Green's
identity
\begin{equation}
\inner{\constop\left(\genpot\right)}{\genpotb}-\inner{\genpot}{\adjoint{\constop}\left(\genpotb\right)}=\bterm\label{eq:Green identity-1}
\end{equation}
where
\begin{equation}
\inner{\constop\left(\genpot\right)}{\genpotb}\coloneqq\intvol{\constop\left(\genpot\right)\cdot\conj{\genpotb}}=\intvol{\left\{ \constop\left(\genpot\right)\right\} _{\alpha}\conj v_{\alpha}},\label{eq:inner product-1}
\end{equation}
and $\bterm$ denotes a surface integral of some bilinear function
of $\genpot,\genpotb$ and their derivatives, where the superscript
$*$ denotes complex conjugate operation. Setting $\genpot=\genpot^{\left(1\right)}$
and $\genpotb=\genpot^{\left(2\right)}$, we obtain \begin{widetext}

\begin{equation}
\begin{aligned} & \inner{\constop\left(\genpot\right)}{\genpotb}=\\
 & \intvol{\left(\left\{ \elasindex_{ijkl}\dispIndex_{k,l}^{(1)}+\tp{\Piezoelectricindex}_{ijk}\potential_{,k}^{(1)}\right\} _{,j}\dispIndex_{i}^{(2)*}+\left\{ \Piezoelectricindex_{ijk}\dispIndex_{k,l}^{(1)}-\Permittivityindex_{ij}\potential_{,j}^{(1)}\right\} _{,i}\potential^{(2)*}+\omega^{2}\rho\dispIndex_{i}^{(1)}\dispIndex_{i}^{(2)*}\right)}=\\
 & \intvol{\left(\dispIndex_{k}^{(1)}\left\{ \elasindex_{ijkl}\dispIndex_{i,j}^{(2)*}+\tp{\Piezoelectricindex}_{kji}\potential_{,i}^{(2)*}\right\} _{,j}+\potential^{(1)}\left\{ \Piezoelectricindex_{ijk}\dispIndex_{j,k}^{(2)*}-\Permittivityindex_{ij}\potential_{,i}^{(2)*}\right\} _{,i}+\dispIndex_{i}^{(1)}\omega^{2}\rho\dispIndex_{i}^{(2)*}\right)}=\\
 & \intvol{\left(\dispIndex_{k}^{(1)}\left\{ \tp{\elasindex}_{klij}\dispIndex_{i,j}^{(2)*}+\tp{\Piezoelectricindex}_{kji}\potential_{,i}^{(2)*}\right\} _{,j}+\potential^{(1)}\left\{ \Piezoelectricindex_{ijk}\dispIndex_{j,k}^{(2)*}-\tp{\Permittivityindex}_{ji}\potential_{,i}^{(2)*}\right\} _{,i}+\dispIndex_{i}^{(1)}\omega^{2}\rho\dispIndex_{i}^{(2)*}\right)}=\\
 & \intvol{\left(\dispIndex_{k}^{(1)}\left\{ \elasindex_{klij}^{\mathsf{T}*}\dispIndex_{i,j}^{(2)}+\Piezoelectricindex_{kji}^{\mathsf{T}*}\potential_{,i}^{(2)}\right\} _{,j}^{*}+\potential^{(1)}\left\{ \Piezoelectricindex_{ijk}^{*}\dispIndex_{j,k}^{(2)}-\Permittivityindex_{ji}^{\mathsf{T}*}\potential_{,i}^{(2)}\right\} _{,i}^{*}+\dispIndex_{i}^{(1)}\left\{ \omega^{2}\rho^{*}\dispIndex_{i}^{(2)}\right\} ^{*}\right)}=\\
 & \inner{\genpot}{\adjoint{\constop}\left(\genpotb\right)},
\end{aligned}
\label{eq:adjoint-appendix}
\end{equation}
\end{widetext}using integration by parts, where the boundary terms
that result in the process are indeed bilinear functions of $\genpot,\genpotb$,
and were omitted from Eq.$\ $\eqref{eq:adjoint-appendix} for brevity.
This identifies $\adjoint{\constop}$ with the adjoint equations
\begin{equation}
\left\{ \adjoint{\genkcomp}_{\alpha i\beta j}\gendiscomp_{\beta,j}\right\} _{,i}+\conj{\rho}\omega^{2}\gend_{\alpha\beta}\gendiscomp_{\beta}=\adjointforce_{\alpha},\label{eq:unified adjoint equations-1}
\end{equation}
where
\begin{equation}
\adjoint{\genkcomp}_{\alpha i\beta j}=\begin{cases}
\elasindex_{\alpha i\beta j}^{\mathsf{T}*} & \alpha,\beta\in\left\{ 1,2,3\right\} ,\\
\Piezoelectricindex_{\alpha i\beta}^{\mathsf{T}*} & \beta=4,\ \alpha\in\left\{ 1,2,3\right\} ,\\
\Piezoelectricscalar_{i\beta j}^{*} & \alpha=4,\ \beta\in\left\{ 1,2,3\right\} ,\\
-\Permittivityscalar_{ij}^{\mathsf{T}*} & \alpha=\beta=4.\\
\end{cases}\label{eq:adj k-1}
\end{equation}
Accordingly, the components of the adjoint green matrix $\adjoint G_{\beta\gamma}\left(\sys,\dummyb\right)$
are defined by
\begin{equation}
\left\{ \adjoint{\genkcomp}_{\alpha i\beta j}\adjoint G_{\beta\gamma,j}\right\} _{,i}+\conj{\rho}\omega^{2}\gend_{\alpha\beta}\adjoint G_{\beta\gamma}=-\delta_{\alpha\gamma}\delta\left(\sys-\dummyb\right).\label{eq:green def-1-1}
\end{equation}
Following the standard procedure, we set $\gendiscomp_{\alpha}\left(\dummyx,\sys\right)=G_{\alpha\left(\gamma\right)}\left(\dummyx,\sys\right)$
and $v_{\alpha}\left(\dummyx,\dummyb\right)=\adjoint G_{\alpha\left(\beta\right)}\left(\dummyx,\dummyb\right)$,
and employ Eqs.$\ $\eqref{eq:unified governing equations}-\eqref{eq:Green identity-1},
\eqref{eq:unified adjoint equations-1} and \eqref{eq:green def-1-1}
to show that\\
\begin{widetext}
\begin{equation}
\begin{aligned} & \inner{\constop\left(\genpot\right)}{\genpotb}-\inner{\genpot}{\adjoint{\constop}\left(\genpotb\right)}=\\
 & \intvoldummy{\left\{ \constop G_{.\left(\gamma\right)}\left(\dummyx,\sys\right)\right\} _{\alpha}G_{\alpha\left(\beta\right)}^{\dagger*}\left(\dummyx,\dummyb\right)}-\intvoldummy{G_{\alpha\left(\gamma\right)}\left(\dummyx,\sys\right)\left\{ \adjoint{\constop}G_{\alpha\left(\beta\right)}^{\dagger}\left(\dummyx,\dummyb\right)\right\} ^{*}}=\\
 & \intvoldummy{\delta_{\alpha\gamma}\delta\left(\dummyx-\sys\right)G_{\alpha\left(\beta\right)}^{\dagger*}\left(\dummyx,\dummyb\right)}-\intvoldummy{G_{\alpha\left(\gamma\right)}\left(\dummyx,\sys\right)\delta_{\alpha\beta}\delta\left(\dummyx-\dummyb\right)}=\\
 & G_{\gamma\left(\beta\right)}^{\dagger*}\left(\sys,\dummyb\right)-G_{\beta\left(\gamma\right)}\left(\dummyb,\sys\right)=0,
\end{aligned}
\label{eq:adjoint-1-1}
\end{equation}
\end{widetext}and hence $G_{\gamma\beta}^{\dagger}\left(\sys,\dummyb\right)=\conj G_{\beta\gamma}\left(\dummyb,\sys\right).$
If the body satisfies $\adjoint{\genkcomp}_{\alpha i\beta j}=\genkcomp_{\alpha i\beta j}$
and $\conj{\rho}=\rho$---which is the case by virtue of Eq.$\ $\eqref{eq:symmetries}---then
the problem is self-adjoint, as Eq.$\ $\eqref{eq:unified governing equations}
is identical to Eq.$\ $\eqref{eq:unified adjoint equations-1}. In
this case $\constop=\adjoint{\constop}$, hence $G_{\gamma\beta}^{\dagger}\left(\sys,\dummyb\right)=G_{\gamma\beta}\left(\sys,\dummyb\right)$,
which together with the previous result implies that
\begin{equation}
G_{\gamma\beta}\left(\sys,\dummyb\right)=\conj G_{\beta\gamma}\left(\dummyb,\sys\right).\label{eq:self adjoint G-1}
\end{equation}

As discussed in Sec.$\ $\ref{sec:Reciprocity}, the symmetries of
$\green$ are required in showing that $\effective{\lmat}$---which
is a function of $\green$---satisfies the symmetries that are given
by Eqs.$\ $\eqref{eq:L self adjoint}-\eqref{eq:Willis type coupling},
\emph{i.e.}, it is self-adjoint. The components of $\effective{\lmat}$
involve lengthy expressions, which we omitted here. We choose however
to provide expression for $\FTrans{\rg}$ and $\adjoint{\FTrans{\rg}}$,
and show the symmetry we reported. We begin by writing the result for $\adjoint{\FTrans{\rg}}$ from Eq.~\eqref{eq:nonlocal L}
as
\begin{eqnarray}
\adjoint{\effective{\rg}}(\position,\dummyb) & = & -\boldsymbol{\alpha}_{32}+\boldsymbol{\gamma}_{32},\label{eq:W}
\end{eqnarray}
where  $\boldsymbol{\alpha}_{32}$ and $\boldsymbol{\gamma}_{32}$ are the $(3,2)$ entries of the symbolic $3\times3$ block matrices
$\upalpha=\ensemble{\lmat\gradt\tp{\left(\gradt\green\right)}\lmat}$
and $\upgamma=\ensemble{\lmat\gradt\green^{\mathsf{T}}}\ensemble{\green}^{-\mathsf{T}}\ensemble{\tp{\left(\gradt\green\right)}\lmat}$, which read

\begin{widetext}
\begin{eqnarray}
\boldsymbol{\alpha}_{32}(\position,\dummyb) & = & s\ensemble{\rho(\position)\left(\Curlp\tp{\greentensor_{11}}\right)\tp{\Piezoelectricmodule}(\dummyb)} -s\ensemble{\rho(\position)\left(\Curlp\tp{\greentensor_{21}}\right)\Permittivity(\dummyb)},\label{eq:alfa23}\\
\boldsymbol{\gamma}_{32}(\position,\position') & = & s\ensemble{\rho(\position)\tp{\greentensor_{11}}}\ensemble{\tp{\Ginv_{11}}}\ensemble{\left(\Curlp\tp{\greentensor_{11}}\right)\tp{\Piezoelectricmodule}(\dummyb)}-s\ensemble{\rho(\position)\tp{\greentensor_{11}}}\ensemble{\tp{\Ginv_{11}}}\ensemble{\left(\Curlp\tp{\greentensor_{21}}\right)\Permittivity(\dummyb)}\nonumber \\
 & + & s\ensemble{\rho(\position)\tp{\greentensor_{21}}}\ensemble{\tp{\Ginv_{12}}}\ensemble{\left(\Curlp\tp{\greentensor_{11}}\right)\tp{\Piezoelectricmodule}(\dummyb)}-s\ensemble{\rho(\position)\tp{\greentensor_{21}}}\ensemble{\tp{\Ginv_{12}}}\ensemble{\left(\Curlp\tp{\greentensor_{21}}\right)\Permittivity(\dummyb)}\nonumber \\
 & + & s\ensemble{\rho(\position)\tp{\greentensor_{11}}}\ensemble{\tp{\Ginv_{21}}}\ensemble{\left(\Curlp\tp{\greentensor_{12}}\right)\tp{\Piezoelectricmodule}(\dummyb)}-s\ensemble{\rho(\position)\tp{\greentensor_{11}}}\ensemble{\tp{\Ginv_{21}}}\ensemble{\left(\Curlp\tp{\greentensor_{22}}\right)\Permittivity(\dummyb)}\nonumber \\
 & + & s\ensemble{\rho(\position)\tp{\greentensor_{21}}}\ensemble{\tp{\Ginv_{22}}}\ensemble{\left(\Curlp\tp{\greentensor_{12}}\right)\tp{\Piezoelectricmodule}(\dummyb)}-s\ensemble{\rho(\position)\tp{\greentensor_{21}}}\ensemble{\tp{\Ginv_{22}}}\ensemble{\left(\Curlp\greentensor_{22}\right)\Permittivity(\dummyb)},\nonumber \\
\label{eq:Gamma23}
\end{eqnarray}
\end{widetext} where
\begin{equation}
\Ginv\equiv\green^{-1}=\left(\begin{array}{cc}
\Ginv_{11} & \Ginv_{12}\\
\Ginv_{21} & \Ginv_{22}
\end{array}\right).
\end{equation}
Note that if $\green$ satisfy the symmetry $\Hadj{\green}(\position,\dummyb)=\green(\dummyb,\position)$
then $\Hadj{\green}(\position,\dummyb)^{-1}=\green(\dummyb,\position)^{-1}$
and consequently, $\Hadj{\Ginv}(\position,\dummyb)=\Ginv(\dummyb,\position)$,
so
\begin{eqnarray}
\Hadj{\Ginv_{\mathrm{\mathit{pq}}}}(\position,\dummyb) & = & \Ginv_{\mathrm{\mathit{qp}}}(\dummyb,\position).\label{eq:V-Sym}
\end{eqnarray}
On the other hand, from Eq.~\eqref{eq:nonlocal L} we have
\begin{equation}
\Hadj{\effective{\rg}}(\dummyb,\position)=-\Hadj{\boldsymbol{\alpha}_{23}}(\dummyb,\position)+\Hadj{\boldsymbol{\gamma}_{23}}(\dummyb,\position),
\end{equation}
where
\begin{widetext}
\begin{eqnarray}
\Hadj{\boldsymbol{\alpha}_{23}}(\dummyb,\position) & =&s^{*}\ensemble{\Hadj{\rho}(\position)\left(\Curlp\greentensor_{11}^{*}\right)\Hadj{\Piezoelectricmodule}(\dummyb)}-s^{*}\ensemble{\Hadj{\rho}(\position)\left(\Curlp\greentensor_{12}^{*}\right)\Hadj{\Permittivity}(\dummyb)},\label{eq:A23Hadj}\\
\Hadj{\boldsymbol{\gamma}_{23}}(\dummyb,\position) & = & s^{*}\ensemble{\Hadj{\rho}(\position)\greentensor_{11}^{*}}\ensemble{\Ginv_{11}^{*}}\ensemble{\left(\Curlp\greentensor_{11}^{*}\right)\Hadj{\Piezoelectricmodule}(\dummyb)}-s^{*}\ensemble{\Hadj{\rho}(\position)\greentensor_{11}^{*}}\ensemble{\Ginv_{11}^{*}}\ensemble{\left(\Curlp\greentensor_{12}^{*}\right)\Hadj{\Permittivity}(\dummyb)}\nonumber \\
 & + & s^{*}\ensemble{\Hadj{\rho}(\position)\greentensor_{11}^{*}}\ensemble{\Ginv_{12}^{*}}\ensemble{\left(\Curlp\greentensor_{21}^{*}\right)\Hadj{\Piezoelectricmodule}(\dummyb)}-s^{*}\ensemble{\Hadj{\rho}(\position)\greentensor_{11}^{*}}\ensemble{\Ginv_{12}^{*}}\ensemble{\left(\Curlp\greentensor_{22}^{*}\right)\Hadj{\Permittivity}(\dummyb)}\nonumber \\
 & + & s^{*}\ensemble{\Hadj{\rho}(\position)\greentensor_{12}^{*}}\ensemble{\Ginv_{21}^{*}}\ensemble{\left(\Curlp\greentensor_{11}^{*}\right)\Hadj{\Piezoelectricmodule}(\dummyb)}-s^{*}\ensemble{\Hadj{\rho}(\position)\greentensor_{12}^{*}}\ensemble{\Ginv_{21}^{*}}\ensemble{\left(\Curlp\greentensor_{12}^{*}\right)\Hadj{\Permittivity}(\dummyb)}\nonumber \\
 & + & s^{*}\ensemble{\Hadj{\rho}(\position)\greentensor_{12}^{*}}\ensemble{\Ginv_{22}^{*}}\ensemble{\left(\Curlp\greentensor_{21}^{*}\right)\Hadj{\Piezoelectricmodule}(\dummyb)}-s^{*}\ensemble{\Hadj{\rho}(\position)\greentensor_{12}^{*}}\ensemble{\Ginv_{22}}\ensemble{\left(\Curlp\greentensor_{22}^{*}\right)\Hadj{\Permittivity}(\dummyb)}.\nonumber \\
\label{eq:eq:Gamma23Hadj}
\end{eqnarray}
\end{widetext} Note that if $\Hadj{\lmat}=\lmat$, since $\Hadj{\green}(\dummyb,\position)=\green(\position,\dummyb)$,
Eq.~\eqref{eq:V-Sym} holds and $\boldsymbol{\alpha}_{32}(\position,\dummyb)=-\Hadj{\boldsymbol{\alpha}_{23}}(\dummyb,\position)$,
 $\boldsymbol{\gamma}_{32}(\position,\dummyb)=-\Hadj{\boldsymbol{\gamma}_{23}}(\dummyb,\position)$,
for $s=-i\omega$, indicating that $\adjoint{\effective{\rg}}(\position,\dummyb)=-\Hadj{\effective{\rg}}(\dummyb,\position)$.
We assume statistically homogeneous media, so $\adjoint{\effective{\rg}}(\position-\dummyb)=-\Hadj{\effective{\rg}}(\dummyb-\position)$
and its Fourier transform leads to the relation $\adjoint{\FTrans{\rg}}(\Wavevector)=-\Hadj{\FTrans{\rg}}(\Wavevector)$
or
\begin{equation}
\adjoint{\FTrans{\rg}}(\Wavevector,\omega)=-\Hadj{\FTrans{\rg}}(\Wavevector,\omega),\label{eq:WHadj-relation}
\end{equation}
which is the first relation in Eq.~\eqref{eq:FTrW}. If $\lmat$ is
real, from the Fourier transform of Eq.~\eqref{eq:L self adjoint}
we have
\begin{equation}
\adjoint{\FTrans{\rg}}(\Wavevector,\omega)=\tp{\FTrans{\rg}}(-\Wavevector,\omega).\label{eq:WAAdj-Relation}
\end{equation}
Finally, from Eqs.~\eqref{eq:WHadj-relation} and~\eqref{eq:WAAdj-Relation}
we obtain the relation
\begin{align}
\FTrans{\rg}(\Wavevector,\omega) & =-\FTrans{\rg}^{*}(-\Wavevector,\omega).
\end{align}

\bibliographystyle{unsrtnat}
\bibliography{bibtexfiletot.bib}

\end{document}